\documentclass[twocolumn, twocolappendix]{aastex63}

\usepackage{ragged2e}   
\usepackage{tabularx}   
\usepackage{tcolorbox}  

\shorttitle{eDisk: R CrA IRAS 32}
\shortauthors{Encalada et al.}

\graphicspath{{./}{figures/}}

\begin{document}

\title{Early Planet Formation in Embedded Disks (eDisk) XIII: Aligned Disks with Non-Settled Dust Around the Newly Resolved Class 0 Protobinary R CrA IRAS 32}

\correspondingauthor{Frankie J. Encalada}
\email{fje2@illinois.edu}

\author[0000-0002-3566-6270]{Frankie J. Encalada}
\affiliation{Department of Astronomy, University of Illinois, 1002 West Green St, Urbana, IL 61801, USA}

\author[0000-0002-4540-6587]{Leslie W. Looney}
\affiliation{Department of Astronomy, University of Illinois, 1002 West Green St, Urbana, IL 61801, USA}


\author[0000-0003-0845-128X]{Shigehisa Takakuwa}
\affiliation{Department of Physics and Astronomy, Graduate School of Science and Engineering, Kagoshima University, 1-21-35 Korimoto, Kagoshima,Kagoshima 890-0065, Japan}
\affiliation{Academia Sinica Institute of Astronomy \& Astrophysics \\
11F of Astronomy-Mathematics Building, AS/NTU, No.1, Sec. 4, Roosevelt Rd \\
Taipei 10617, Taiwan, R.O.C.}

\author[0000-0002-6195-0152]{John J. Tobin}
\affil{National Radio Astronomy Observatory, 520 Edgemont Rd., Charlottesville, VA 22903 USA}

\author[0000-0003-0998-5064]{Nagayoshi Ohashi}
\affiliation{Academia Sinica Institute of Astronomy \& Astrophysics \\
11F of Astronomy-Mathematics Building, AS/NTU, No.1, Sec. 4, Roosevelt Rd \\
Taipei 10617, Taiwan, R.O.C.}

\author[0000-0001-9133-8047]{Jes K. J{\o}rgensen}
\affil{Niels Bohr Institute, University of Copenhagen, {\O}ster Voldgade 5--7, DK~1350 Copenhagen K., Denmark}

\author[0000-0002-7402-6487]{Zhi-Yun Li}
\affiliation{University of Virginia, 530 McCormick Rd., Charlottesville, Virginia 22904, USA}


\author[0000-0003-3283-6884]{Yuri Aikawa}
\affiliation{Department of Astronomy, Graduate School of Science, The University of Tokyo, 7-3-1 Hongo, Bunkyo-ku, Tokyo 113-0033, Japan}

\author[0000-0002-8238-7709]{Yusuke Aso}
\affiliation{Korea Astronomy and Space Science Institute, 776 Daedeok-daero, Yuseong-gu, Daejeon 34055, Republic of Korea}

\author[0000-0003-2777-5861]{Patrick M. Koch}
\affiliation{Academia Sinica Institute of Astronomy \& Astrophysics \\
11F of Astronomy-Mathematics Building, AS/NTU, No.1, Sec. 4, Roosevelt Rd \\
Taipei 10617, Taiwan, R.O.C.}

\author[0000-0003-4022-4132]{Woojin Kwon}
\affiliation{Department of Earth Science Education, Seoul National University, 1 Gwanak-ro, Gwanak-gu, Seoul 08826, Republic of Korea}
\affiliation{SNU Astronomy Research Center, Seoul National University, 1 Gwanak-ro, Gwanak-gu, Seoul 08826, Republic of Korea}

\author[0000-0001-5522-486X]{Shih-Ping Lai}
\affiliation{Institute of Astronomy, National Tsing Hua University, No. 101, Section 2, Kuang-Fu Road, Hsinchu 30013, Taiwan}
\affiliation{Center for Informatics and Computation in Astronomy, National Tsing Hua University, No. 101, Section 2, Kuang-Fu Road, Hsinchu 30013, Taiwan}
\affiliation{Department of Physics, National Tsing Hua University, No. 101, Section 2, Kuang-Fu Road, Hsinchu 30013, Taiwan}
\affiliation{Academia Sinica Institute of Astronomy \& Astrophysics \\
11F of Astronomy-Mathematics Building, AS/NTU, No.1, Sec. 4, Roosevelt Rd \\
Taipei 10617, Taiwan, R.O.C.}

\author[0000-0002-3179-6334]{Chang Won Lee}
\affiliation{Korea Astronomy and Space Science Institute, 776 Daedeok-daero Yuseong-gu, Daejeon 34055, Republic of Korea}
\affiliation{University of Science and Technology, 217 Gajeong-ro Yuseong-gu, Daejeon 34113, Republic of Korea}

\author[0000-0001-7233-4171]{Zhe-Yu Daniel Lin}
\affiliation{University of Virginia, 530 McCormick Rd., Charlottesville, Virginia 22904, USA}

\author[0000-0001-6267-2820]{Alejandro Santamaría-Miranda}
\affiliation{European Southern Observatory, Alonso de Cordova 3107, Casilla 19, Vitacura, Santiago, Chile}

\author[0000-0003-4518-407X]{Itziar de Gregorio-Monsalvo}
\affiliation{European Southern Observatory, Alonso de Cordova 3107, Casilla 19, Vitacura, Santiago, Chile}

\author[0000-0002-4372-5509]{Nguyen Thi Phuong}
\affiliation{Korea Astronomy and Space Science Institute, 776 Daedeok-daero, Yuseong-gu, Daejeon 34055, Republic of Korea\\}
\affiliation{Department of Astrophysics, Vietnam National Space Center, Vietnam Academy of Science and Techonology, 18 Hoang Quoc Viet, Cau Giay, Hanoi, Vietnam}

\author[0000-0002-9912-5705]{Adele Plunkett}
\affiliation{National Radio Astronomy Observatory, 520 Edgemont Rd., Charlottesville, VA 22903 USA}

\author[0000-0003-4361-5577]{Jinshi Sai (Insa Choi)}
\affiliation{Academia Sinica Institute of Astronomy \& Astrophysics \\
11F of Astronomy-Mathematics Building, AS/NTU, No.1, Sec. 4, Roosevelt Rd \\
Taipei 10617, Taiwan, R.O.C.}

\author[0000-0002-0549-544X]{Rajeeb Sharma}
\affiliation{Niels Bohr Institute, University of Copenhagen, \O ster Voldgade 5--7, 1350, Copenhagen K, Denmark}

\author[0000-0003-1412-893X]{Hsi-Wei Yen}
\affiliation{Academia Sinica Institute of Astronomy \& Astrophysics \\
11F of Astronomy-Mathematics Building, AS/NTU, No.1, Sec. 4, Roosevelt Rd \\
Taipei 10617, Taiwan, R.O.C.}

\author[0000-0002-9143-1433]{Ilseung Han}
\affiliation{Division of Astronomy and Space Science, University of Science and Technology, 217 Gajeong-ro, Yuseong-gu, Daejeon 34113, Republic of Korea}
\affiliation{Korea Astronomy and Space Science Institute, 776 Daedeok-daero, Yuseong-gu, Daejeon 34055, Republic of Korea}



\begin{abstract}

Young protostellar binary systems, with expected ages less than $\sim$10$^5$ years, are little modified since birth, providing key clues to binary formation and evolution. We present a first look at the young, Class 0 binary protostellar system R CrA IRAS 32 from the Early Planet Formation in Embedded Disks (eDisk) ALMA large program, which observed the system in the 1.3 mm continuum emission, $^{12}$CO (2-1), $^{13}$CO (2-1), C$^{18}$O (2-1), SO (6$_5$-5$_4$), and nine other molecular lines that trace disk, envelope, shocks, and outflows. With a continuum resolution of $\sim$0.03$^{\prime\prime}$ ($\sim$5 au, at a distance of 150 pc), we characterize the newly discovered binary system with a separation of 207 au, their circumstellar disks, and a circumbinary disk-like structure. The circumstellar disk radii
are 26.9$\pm$0.3 and 22.8$\pm$0.3 au for sources A and B, respectively, and their circumstellar disk dust masses are estimated as 22.5$\pm$1.1 and 12.4$\pm$0.6 M$_{\Earth}$. The circumstellar disks and the circumbinary structure have well aligned position angles and inclinations, indicating formation in a smooth, ordered process such as disk fragmentation. In addition, the circumstellar disks have a near/far-side asymmetry in the continuum emission suggesting that the dust has yet to settle into a thin layer near the midplane. Spectral analysis of CO isotopologues reveals outflows that originate from both of the sources and possibly from the circumbinary disk-like structure. Furthermore, we detect Keplerian rotation in the $^{13}$CO isotopologues toward both circumstellar disks and likely Keplerian rotation in the circumbinary structure; the latter suggests that it is probably a circumbinary disk.

\end{abstract}




\section{Introduction} \label{sec:intro}
The circumstellar disks around young stellar objects are the mass reservoirs from which planets are eventually formed. The structures and properties of the disk, especially in the early stages of protostellar evolution, are critical for understanding star and planet formation. The youngest disks, around the so-called Class 0 sources \citep{andre1993}, are accreting mass and angular momentum from the surrounding envelope while also losing angular momentum via low-velocity outflows and high-velocity jets \citep{snell1980,frank2014}. 
Class 0 disks are critical for star and planet evolution as they are the conduit for envelope accretion onto the protostar, and they are the site for dust evolution and the beginning of planetesimals \citep[e.g.,][]{Testi2014}.

Current observations reveal that gaps and rings are ubiquitous in Class II protostellar disks \citep[e.g.,][]{brogan2015,andrews2018}.  The substructures in the disks are likely caused by planetary-mass bodies in the gaps, shepherding the dust into sharply defined rings or spiral structures \citep{dong2015,zhang2018,huang2018}.  The idea that protoplanets are playing a role in the substructure formation has also been supported by localized gas deviations from Keplerian rotation that are consistent with protoplanets in the gaps \citep{Pinte2020}. 

Although protoplanets creating these substructures appear to be the most likely creation mechanism, 
other possible explanations for the gaps and rings have been suggested, such as 1) non-ideal magnetohydrodynamics effects, e.g., ambipolar diffusion on $\sim$10 au scales with a magnetically coupled disk-wind \citep{suriano2018}, 2) gas density variations at the outer edge of the dead-zone in a magnetorotational instability driven accretion disk \citep{ruge2015}, 3) radial variations in dust properties due to volatiles freezing onto dust grains (i.e., snowlines) \citep{zhang2015}, or 4)  heterogenous infall onto the disk \citep[e.g.,][]{kuznetsova2022}, among others.
In any formation process, the gaps and rings (and other substructures such as spirals) will still have local pressure deviations that trap dust grains, which then promote planetesimal formation \citep[e.g.,][]{vanderMarel2013,perez2016}, so the evolution of substructures is still likely a fundamental step in understanding planet formation regardless of their origins.

It is now clear that a large fraction of T Tauri stars have clear and well-developed substructures \citep{Bae2022}. Many theoretical studies have suggested that if such structures are produced by already-formed planets, planet formation should have started at an earlier time when there was a large reservoir of material \citep[e.g.,][]{tsukamoto2017}.  However, observing Class 0/I objects is more difficult due to the disks still being deeply embedded in an envelope. Nonetheless, recent observations have suggested that substructures can form early in the disk formation process in some cases. \cite{sheehan2020} showed substructure in seven sources of the VLA/ALMA Nascent Disk and Multiplicity Survey of Orion Protostars  \citep{tobin2020}. Three of the sources were modeled to have a very large envelope-to-disk masses, suggesting very young disks.  In addition, the Class I source Oph IRS63 was shown to have multiple narrow (and shallow) annular substructures \citep{dom2020}.
 
To probe the youngest protostellar disk for substructures, measure young disk properties, observe the outflows, probe the disk kinematics, and constrain the protostellar mass, the Atacama Large Millimeter/submillimeter Array (ALMA) Large Program, Early Planet Formation in Embedded Disks \citep[eDisk,][]{edisk}, focuses on the embedded disks in Class 0/I systems. In this program, 17 YSOs in nearby star forming regions were observed, and 2 YSOs were used from the archive, at 1.3 mm (230 GHz) to quantify the kinematics and structures of the embedded disks. The main means of this analysis is through the continuum and CO isotopologues, with a dozen other complementary molecular lines that probe other aspects of the protostellar system.

\begin{table*}[t!]
    \centering
    \caption{Observation Information} \label{tab:obs_data}
    \begin{tabular}{c c c c c c c c c} 
\hline \hline
Observation & Time On       & Array         & Number of & Baselines & PWV$^{c}$   & \multicolumn{3}{c}{Calibrators}       \\ 
Date         & Source       & Config.$^{a}$ & Antennas$^{b}$  &           &       & Bandpass      & Phase         & Flux  \\
(yyyy-mm-dd) & (mm:ss.ss)   &         &           & (m)       & (mm)  &               &               &       \\

\hline

 2021-05-04  & 13:05    & C4/5 & 42        & 15 - 2500       & 1.72 & J1924-2914    & J1937-3958    & J1839-3453\\
 
 2021-05-15  & 13:05    & C6   & 45        & 15 - 2500  & 1.01 & J1924-2914    & J1937-3958   & J1839-3453\\
 
 2021-08-18  & 21:31    & C8   & 40        & 92 - 8300  & 0.33 & J1924-2914    & J1839-3453   & J1823-3454\\
 
 2021-08-22  & 21:31    & C8   & 49        & 47 - 11600 & 0.30 & J1924-2914    & J1839-3453   & J1823-3454\\
 
 2021-10-01  & 21:31    & C8      & 47        & 70 - 11600 & 1.55 & J1924-2914    & J1839-3453   & J1823-3454\\
 
 2021-10-02  & 21:31    & C8      & 43        & 70 - 11600 & 0.73 & J1924-2914    & J1839-3453   & J1826-3650\\

\hline
\end{tabular}
\tablenotetext{}{$^{a}$As reported in the ALMA Scheduling Block. $^{b}$ Maximum number of antennas available during observation.
$^{c}$Taken as the average over the on-source observations.}
\end{table*}

The youngest multiple systems, of which eDisk was found to contain 4, are important to examine in order to reveal their possible formation mechanisms.\citep{edisk}. Although main sequence stars have a multiplicity rate that depends on stellar mass \citep[increasing with mass, e.g.,][]{raghavan2010,sana2011,duchene2013,ward-duong2015,moe2017},
it has been shown that the multiplicity fraction appears to be the largest among the youngest protostars with observed peaks at separations of $\sim$100~au and $\sim$1000~au \citep{tobin2016,encalada2021,tobin2022}. These separations provide evidence of both (1) disk fragmentation by gravitational instability and turbulent fragmentation with migration for systems $<$500 au and of (2) turbulent fragmentation for systems $>$1000~au \citep[e.g.,][]{adams1989,lee2019,padoan2002}.  The youngest protostars are not only setting the stage for the growth of planetesimals in the young disk, but also for multiplicity.

In this paper, we will present the eDisk observations of the young protobinary system R CrA IRAS 32 (IRAS 18595-3712), hereafter IRAS 32.  These are the first observations to resolve the binary.
The Corona Australis dark cloud complex is a nearby star-forming region in the southern skies. It is situated toward the galactic center but well away from the galactic plane ($l$=359.84$^\circ$, $b$=-18.12$^\circ$). IRAS 32 was classified as a Class I protostellar system \citep{wilkings1992} toward the Rossano Cloud B in Corona Australis \citep{rossano1978, nutter2005, bresnahan2018}. However, a more careful fit to the photometry of the source gives a bolometric temperature of 64 K, which suggests that IRAS 32 is better described as a Class 0 source \citep{edisk}.
IRAS 32 has been included in both Herschel and Spitzer surveys of the Corona Australis cloud \citep[e.g.,][]{bresnahan2018,peterson2011} with a known large NE-SW outflow cavity \citep{seale2008}.  
The distance is taken to be 150 pc as a convenient value that is in agreement with recent Gaia parallax estimations \citep{bracco2020,galli2020,zucker2020,dzib2018}.

The paper is organized as follows: in section 2 we present the observations and data reduction, section 3 covers the results of the data reduction, section 4 discusses the results, and section 5 contains our conclusions.

\section{Observations and Data Reduction} \label{sec:observations}
IRAS 32 was observed as part of eDisk, ALMA Large Cycle 7 Program, project 2019.1.00261.L (PI N. Ohashi) in Band 6 (230 GHz, 1.3 mm). 
Here we briefly summarize the observations; see \cite{edisk} for more details. As described in Table \ref{tab:obs_data}, data from  single pointings with only the 12 m array were collected between 2021 May 4th and 2021 October 2nd in six execution blocks, which included the eDisk source IRS5N. The blocks individually lasted either 52 or 81 minutes, corresponding to the two short baseline and four long baseline observations. The two short baseline blocks were in the C4/5 and C6 configuration, with the antenna separations ranging between 15 m and 2500 m;  the shortest baselines set the maximum recoverable scale for the data to 4.1$^{\prime\prime}$. 
The four long baseline blocks (all C8) had the antenna array elements at distances between 47 m and 11600 m.

The products were 1.3 mm continuum along with a slew of molecular lines ($^{12}$CO, $^{13}$CO, C$^{18}$O, SO, H$_{2}$CO, c-C$_{3}$H$_{2}$, CH$_{3}$OH, DCN, and SiO), see Table \ref{tab:spec_data} for more details. 
Due to the large time difference between the scheduling blocks and dynamic scheduling, different sources were used as calibrators, see Table \ref{tab:obs_data} and \cite{edisk} for specifics.

As per the ALMA handbook, the absolute flux calibration accuracy is assumed to be $\sim$10\%. For the remainder of this paper any uncertainties are only statistical in nature.

All six of the ALMA calibrated data sets were combined to create the final product. They were imaged within CASA, Common Astronomy Software Applications \citep{casa2022}, version 6.2.1.7. All execution blocks had their flux rescaled to be in line with the execution block from 2021 August 18th. 

We performed continuum self-calibration on the short baseline observations alone, then separately on the combined short and long baseline observations with different thresholds and solution intervals, and then compared the corrected data to verify that the results were similar for the short baselines.  For the final combined solutions, initial maps were created to assess the clean thresholds per solution interval.
We did four iterations of phase-only self-calibration at solution intervals of \texttt{inf-EB}, \texttt{inf}, 18.14 sec, and \texttt{int}.
A solution interval of \texttt{inf-EB} encompasses the whole execution block of the observation (when combined with \texttt{combine='scan,spw'} during the \texttt{gaincal} step of applying the derived gains to the visibilities), which corrects time invariant phase errors in the data.
The solution interval \texttt{inf} encompasses each scan (when combined with \texttt{combine='spw'}), which begins to correct time variant phase errors. The solution interval of \texttt{int} uses every integration, which corrects time variant phase errors down to the integration time. The time intervals are iteratively applied to the data, building the final solutions.
Subsequently, amplitude and phase self-calibration on the continuum was done with an interval of \texttt{inf}, since we had high S/N and the dynamic range was still limited after phase-only self-calibration by amplitude errors, further iterations showed no improvements. 

We used Briggs weighting with a robust parameter of -0.5 for all the final images since it slightly favors uniform weighting over natural weighting. The deconvolver was \texttt{mtmfs}, or Multi-Term Multi-Frequency Synthesis \citep{rau2011}. 
All maps were primary beam corrected.

\begin{table*}[t]
    \centering
    \caption{Spectral Window Information} \label{tab:spec_data}
    \begin{tabular}{l c c c c c c} 
\hline \hline
Molecule        & Transition                     & Start Freq.$^a$   & End Freq.$^a$     & $\nu_{rest}$  & $\Delta\nu$   & $\Delta$v     \\ 
                &                       & (GHz)         & (GHz)         & (GHz)         & (kHz)         & (km s$^{-1}$) \\
\hline
SiO             & 5-4                   & 217.032808    & 217.177398    & 217.104980    & 970.407       & 1.35          \\
DCN             & 3-2                   & 217.224484    & 217.245846    & 217.238600    & 971.004       & 1.35          \\
$c$-C$_3$H$_2$$^b$& 6$_{0,6}$-5$_{1,5}$   & 217.808970    & 217.830389    & 217.822036    & 973.612       & 1.34          \\
$c$-C$_3$H$_2$$^b$& 6$_{1,6}$-5$_{0,5}$   & 217.808970    & 217.830389    & 217.822150    & 973.612       & 1.34          \\
$c$-C$_3$H$_2$    & 5$_{1,4}$-4$_{2,3}$   & 217.926863    & 217.948294    & 217.940050    & 974.139       & 1.34          \\
$c$-C$_3$H$_2$    & 5$_{2,4}$-4$_{1,3}$   & 218.147239    & 218.168692    & 218.160440    & 975.125       & 1.34          \\
H$_{2}$CO       & 3$_{0,3}$-2$_{0,2}$   & 218.208012    & 218.229471    & 218.222192    & 975.401       & 1.34          \\
CH$_3$OH        & 4$_2$-3$_1E$          & 218.425869    & 218.447349    & 218.440063    & 976.374       & 1.34          \\
H$_{2}$CO       & 3$_{2,2}$-2$_{2,1}$   & 218.461436    & 218.482920    & 218.475632    & 976.533       & 1.34          \\
H$_{2}$CO       & 3$_{2,1}$-2$_{2,0}$   & 218.749578    & 218.764079    & 218.760066    & 121.861       & 0.17          \\
C$^{18}$O       & 2-1                   & 219.549828    & 219.564382    & 219.560354    & 122.307       & 0.17          \\
SO              & 6$_{5}$-5$_{4}$       & 219.938897    & 219.953477    & 219.949433    & 122.523       & 0.17          \\
$^{13}$CO       & 2-1                   & 220.388118    & 220.402728    & 220.398684    & 122.774       & 0.17          \\
$^{12}$CO       & 2-1                   & 230.461570    & 230.614899    & 230.538000    & 488.310       & 0.63          \\

\hline
\end{tabular}
\tablenotetext{}{$^a$ Frequency coverage of the spectral windows that contain the line. $^b$ Blended molecular lines.}
\end{table*}

The spectral windows were continuum subtracted, imaged, and used to produce data cubes. The imaging process followed mostly the same steps as with the continuum. However, the masking was done automatically using \texttt{auto-multithresh} \citep{kepley2020} with the default parameters over the ranges of spectral windows containing specific molecular lines instead of over the full window. We also used \texttt{briggsbwtaper} weighting, which is the standard Briggs weighting but with an inverse uv taper added per channel, with a robust parameter of 0.5 for all the final molecular line images. 
Specific spectral line observational details are in Table \ref{tab:spec_data}. All the eDisk data products will soon be available within the ALMA science archive, see \cite{edisk} for more details.



\section{Results} \label{sec:results}
\subsection{Continuum data}
\begin{figure*}
  \centering
    \includegraphics[width=0.90\textwidth]{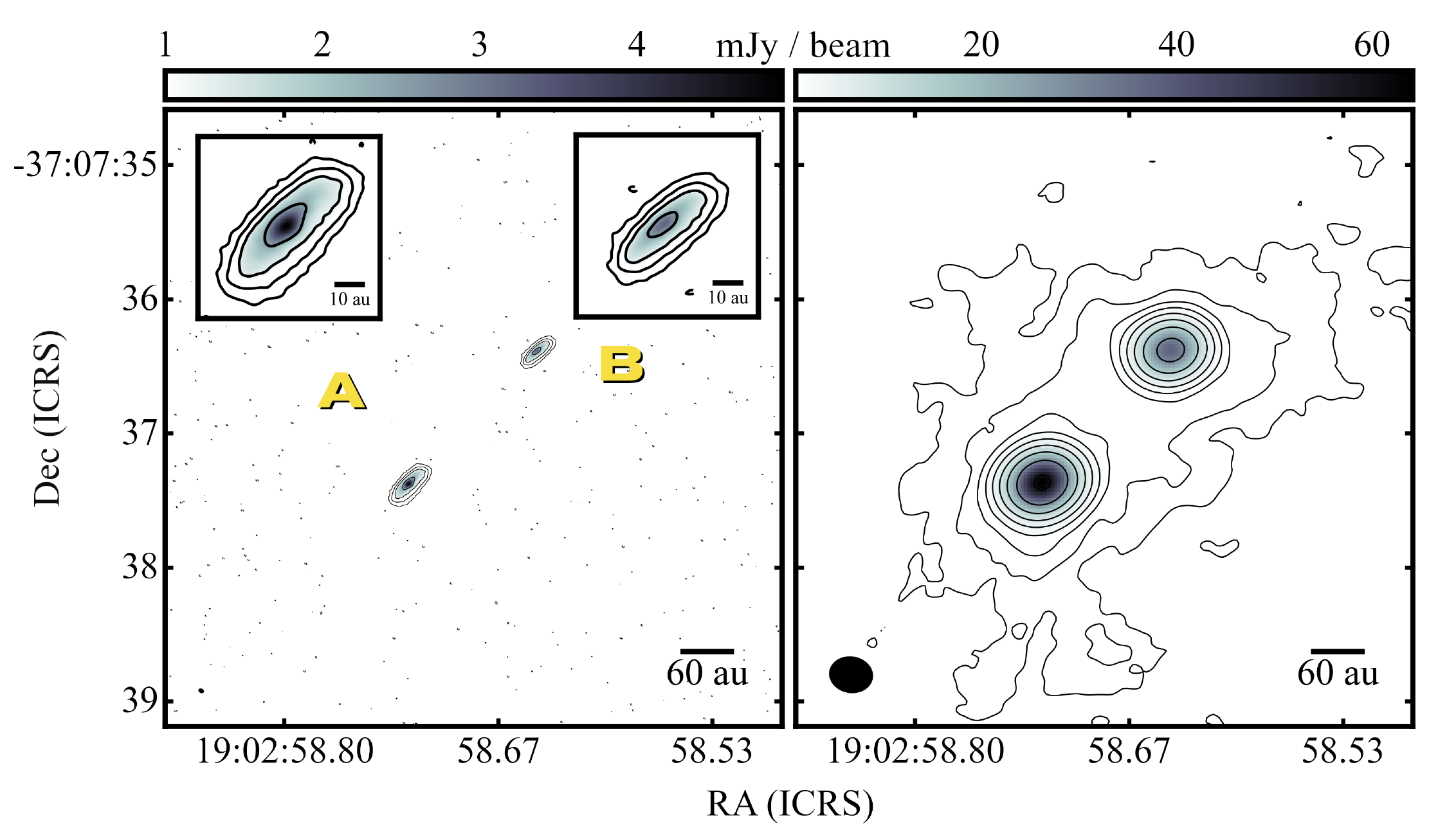}
  \caption{ALMA 1.3 mm continuum maps of IRAS 32. Source A is to the south-east and source B is to the north-west. The beam in the bottom left corner is shown in black. The 60 au scale bar in the bottom right corner represents four-tenths of an arcsec. {\bf Left:} Long and short baselines combined with a Briggs robust parameter of -0.5. Zoomed in views of the sources are in the upper boxes, with source A on the left and source B on the right. Contour levels are shown at -3$\sigma$ and 3$\sigma$2$^{n}$ for $\sigma$=0.0256 mJy beam$^{-1}$ and n$\in$[1,3,4,5]. The beam has a major and minor axis of 0$\farcs$035 $\times$ 0$\farcs$023 and a position angle of 66.6$^{\circ}$. {\bf Right:} Short baseline configuration only with a Briggs robust parameter of 2.0, which favors natural weighting. Contour levels are shown at 3$\sigma$2$^{n}$ for $\sigma$=0.0764 mJy beam$^{-1}$ and n$\in\mathbb{N}_{0}$. The beam has a major and minor axis of 0$\farcs$319  $\times$ 0$\farcs$264, respectively, with a position angle of 80.9$^{\circ}$.}
  \label{fig:900}
\end{figure*}

Fig. \ref{fig:900} shows the 1.3 mm (225 GHz) continuum map of IRAS 32. 
The source is resolved into a binary for the first time, with the A component in the southeast and the B component in the northwest. 
The binary has a projected separation of 1.38$^{\prime\prime}$ (207 au) and a separation angle of 135.9$^{\circ}$ (East of North), with the A component containing nearly twice as much flux density as the B component. 
In addition, when using only the short baseline data of the project, we can make a low resolution map (Fig. \ref{fig:900}, right) that shows a clear flattened circumbinary dust structure surrounding the two disks.
We can derive disk properties directly from the images and from simple Gaussian fits to the two disks and the circumbinary structure.


The peak intensities for the A and B sources are taken from the emission map (Fig. \ref{fig:900}, left) as 4.92 mJy beam$^{-1}$ (148 K) and 3.21 mJy beam$^{-1}$ (96 K), respectively, where the RMS noise is 0.02 mJy beam$^{-1}$. 
We do a two Gaussian fit to Fig. \ref{fig:900}, left, for each disk simultaneously to measure the flux density (F$_{\nu}$) and the deconvolved minor/major axes ($a$ and $b$, respectively) and PA of the disks, as listed in Table \ref{tab:cont_data}. The Gaussian fit peak intensity is also given but not used anywhere in the analysis since the fit peak is less robust than the total flux.
 We do another simultaneous three Gaussian fit to the low resolution map (Fig. \ref{fig:900}, right, two for each disk and one for the circumbinary structure) to measure the circumbinary dust parameters, which is also listed in Table \ref{tab:cont_data}.
Note that the two disk PAs are aligned within a margin of a few degrees, as well as the PA of the overall circumbinary disk structure (see Table \ref{tab:cont_data}).

\begin{table*}
    \centering
    \caption{Gaussian-fit Continuum Information} \label{tab:cont_data}
    \begin{tabularx}{0.86\textwidth}{l c c c c c c c c} 
\hline \hline
Source      & RA            & Dec           & F$_{\nu}$         & Peak I$_{\nu}$    & a     & b     & PA            \\ 
            & (ICRS)        & (ICRS)        & (mJy)             & (mJy bm$^{-1}$)   & (mas) & (mas) & ($^{\circ}$)  \\
            
\hline

 IRAS 32 A  & 19:02:58.7   & -37:07:37.38  & 79.7$\pm$0.9	   & 4.92$\pm$0.02     & 212$\pm$2 & 78$\pm$1  & 135.3$\pm$0.4 \\
 IRAS 32 B  & 19:02:58.6   & -37:07:36.39  & 43.8$\pm$0.4	   & 3.21$\pm$0.02     & 180$\pm$2 & 57.9$\pm$0.7  & 131.5$\pm$0.3 \\
 Circumbinary$^{a}$ & 19:02:58.7    &   -37:07:36.88           & 46                & 0.5               & 3831          & 1903          & 140.7                      \\


\hline
\end{tabularx}

\footnotesize{The beam is 35 $\times$ 23 mas with a PA of 66.6$\degr$. The RMS is 0.026 mJy beam$^{-1}$. $^{a}$ The short baseline only continuum map was also used to simultaneously fit the circumbinary structure with the two compact sources. The main objective of this fit was to quantify the circumbinary structure PA. The other circumbinary structure quantities are also given but the errors should be assumed to be of order $\sim$10\%. We give the binary midpoint as the RA and DEC, as this location is used as the center of some maps.}

\end{table*}

In Table \ref{tab:cont_derived_data}, we include the derived quantities of the continuum data presented in Table \ref{tab:cont_data}. The inclination, $i$, is estimated by using $\arccos{\frac{b}{a}}$, where a and b are the major and minor axes of the fitted Gaussian. The radius ($R_{\rm disk}$) is taken as the 2$\sigma$ value (FWHM/$\sqrt{2{\rm ln2}}$) of the Gaussian-fit deconvolved major axis. The dust mass estimate assumes isothermal and optically thin emission and is derived from

\begin{equation}
M_{dust} = \frac{d^{2} F_{\nu}}{B_{\nu}(T_{dust}) \kappa_{\nu}}
\label{eq:mdust}
\end{equation}

\noindent where $d$ is the distance to the source (150 pc), $F_{\rm \nu}$ is the  observed flux density, and $B_{\rm \nu}$ is the Planck function. We adopt a dust mass opacity, $\kappa_{\nu}$ = 2.3~cm$^{2}$ g$^{-1}$ at $\nu=225$~GHz \citep{beckwith1990}.

We use two temperatures for the mass estimate. First, we adopt a temperature of 20 K. In that case, we calculate the dust masses of the two disks as
M$_{\rm dust,A,20K}$ = 53.0$\pm$2.6 M$_{\Earth}$ and M$_{\rm  dust,B,20K}$ = 29.1$\pm$1.4 M$_{\Earth}$.
In the second case, we estimate the temperature, $T_{\rm  dust}$, given the bolometric luminosity prescription from \cite{tobin2020}

\begin{equation}
T_{dust} = T_{0} \left( \frac{L_{bol}}{1 ~L_{\odot}} \right) ^{0.25},
\label{eq:tdust}
\end{equation}

\noindent with $T_{0}$ = 43 K. IRAS 32 has L$_{bol}$ = 1.6 L$_{\odot}$ via a spectral energy distribution (SED) fit compiled from the NASA/IPAC Infrared Science Archive (IRSA) database \citep[see][]{edisk}. 
It is not possible to accurately estimate the individual protostellar contributions to the total luminosity, but due to their similarities, we assign 50\% of the luminosity to each of them, which
gives T$_{dust}$=40.7 K.
Putting it all together, our derived dust masses are M$_{dust,A}$ = 22.5$\pm$1.1 M$_{\Earth}$ and M$_{dust,B}$ = 12.4$\pm$0.6 M$_{\Earth}$.  As the luminosity prescription gives nearly twice the temperature, it estimates a lower disk mass. 
It should also be mentioned that the IRAS 32 disks are somewhat smaller in radii than those used to derive the 43 K normalization in \cite{tobin2020} (50 au), so we would expect our disks to actually be even warmer on average to this estimate. In addition, since the disks are likely not optically thin, the estimates here are disk mass lower limits.

\begin{center}
\begin{table}[t]
    \caption{Derived Continuum Information} \label{tab:cont_derived_data}
    \begin{center}
\begin{tabular}{l c c c c c} 
\hline \hline
Source  & i             & M$_{dust,40.7K}$$^{a}$ & M$_{dust,20K}$$^{b}$   & R$_{disk}$     \\ 
        & ($^{\circ}$)  & (M$_{\Earth}$)    & (M$_{\Earth}$)    & (au)  \\            
\hline
 A      & 69            & 23$\pm$1      & 53$\pm$3      & 26.9$\pm$0.3  \\
 B      & 71            & 12.4$\pm$0.6      & 29$\pm$1      & 22.8$\pm$0.3  \\

\hline
\end{tabular}
\end{center}

\tablecomments{Secondary parameters derived from the Gaussian fit of the 1.3 mm emission shown in Table \ref{tab:cont_data}. 
Dust mass assumes a $\kappa_{\nu}$ of 2.3 cm$^{2}$ g$^{-1}$. 
$^{a}$The dust mass derived using a temperature of 40.7K from 
Eq. \ref{eq:tdust}.
$^{b}$The dust mass derived using an assumed temperature of 20K.} 
\end{table}
\end{center}

\subsubsection{Radial profile}

\begin{figure}
  \raggedleft
    \includegraphics[width=0.48\textwidth]{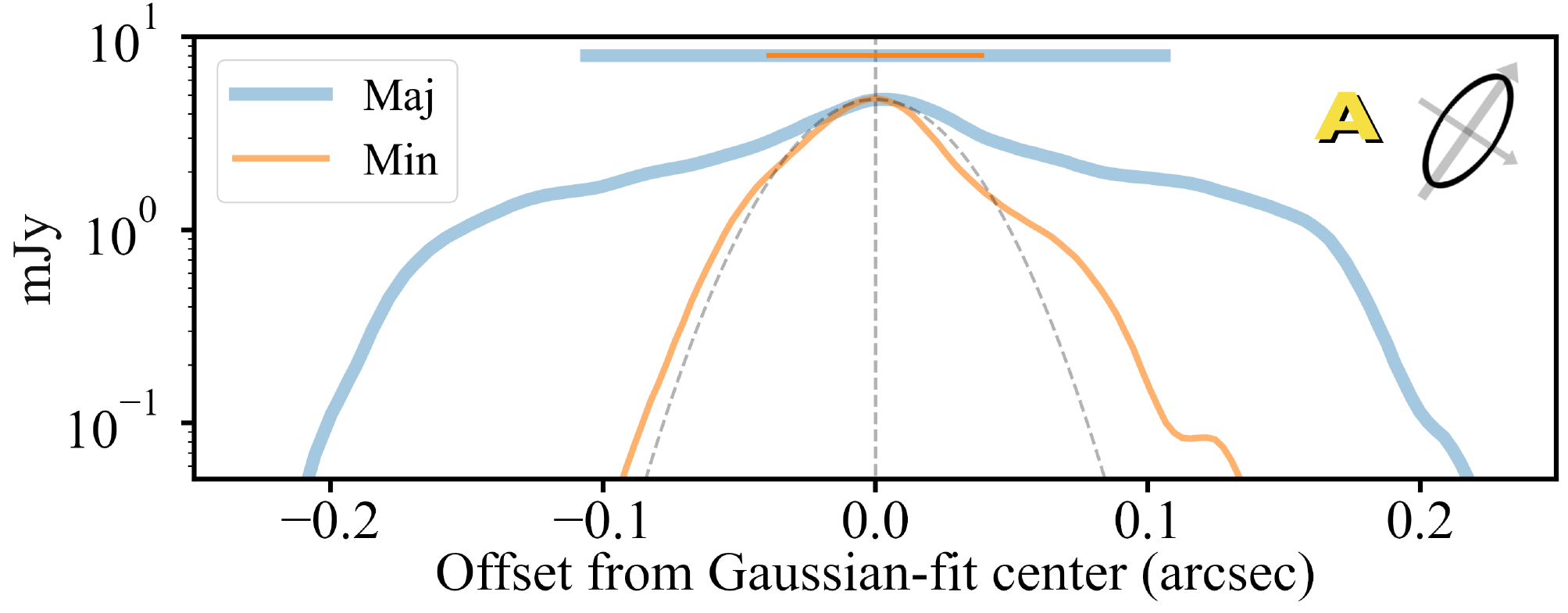}
    \includegraphics[width=0.48\textwidth]{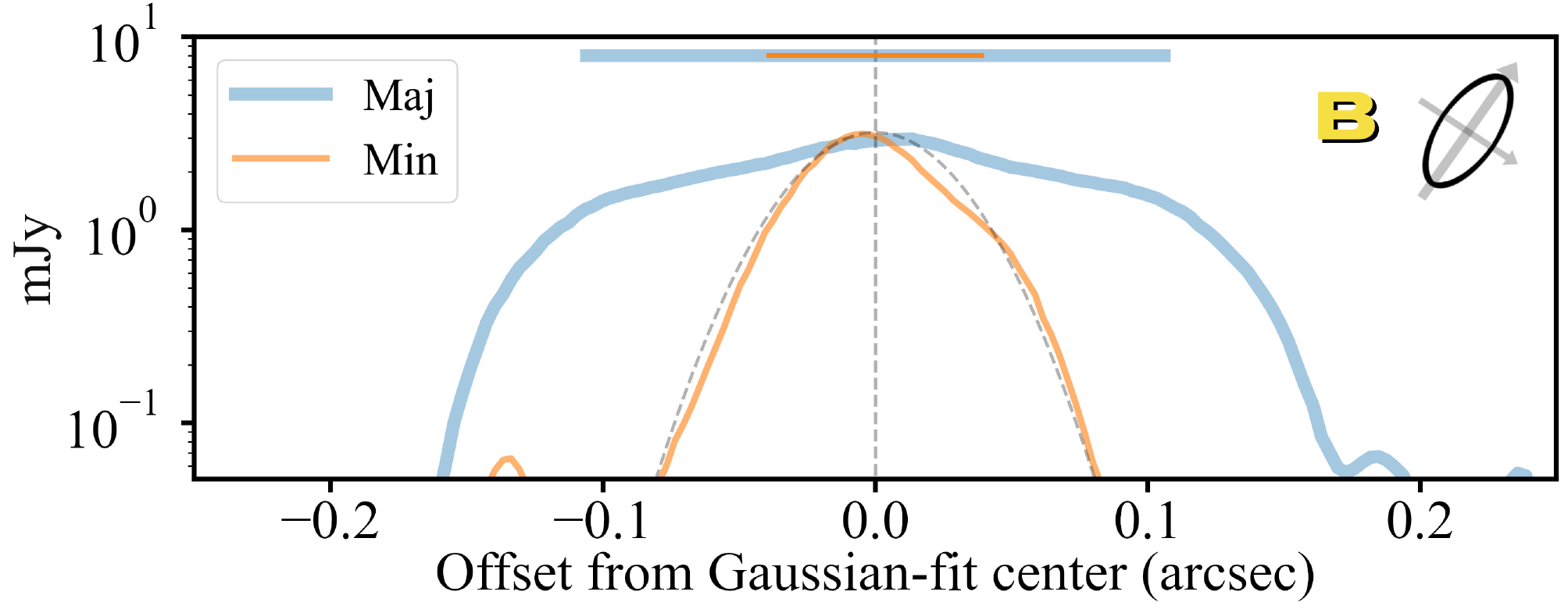}
  \caption{IRAS 32 continuum line projections for both sources. The y-axis is in logarithmic scale. The line projections for both sources are along the major and minor axis and passing through the Gaussian-fit center. The direction of travel is always from east to west. The thickness of the continuum line projection is 0.02$^{\prime\prime}$. A single small diagram showing the direction of projection on the right is used to represent both sources since their PAs are similar. The dashed gray curved line is an simple Gaussian for comparison. Corresponding horizontal lines at the top of the images refer to the Gaussian-fit parameters $a$ and $b$ for their respective sources. The bottom of the y-axis is set at 2$\sigma$ of the RMS.}
  \label{fig:905c}
\end{figure}

While Gaussian-fitting is used to derive basic parameters and quantify the structure of our continuum data, we must ask how accurate this assumption is. In Fig. \ref{fig:905c}, we show intensity profiles of the 1.3-mm continuum emission (from Fig. \ref{fig:900} left) from  along the major and minor axis of both sources, centered on the Gaussian-fit peaks. The separation between the sources gives a high degree of confidence that there is no overlapping contamination from the other source. The first feature we notice is the non-Gaussian nature of the major axis for both sources on both sides of the center. On the other hand, the same can only be said for both sources in the western side of the minor axis. The second noticeable feature is the shift in peak position from the Gaussian-fit center location. This offset can also be visually confirmed in Fig. \ref{fig:900}, left. We will discuss this further in Section \ref{cont-asym}.

\subsection{Molecular line data}
For a full overview of the spectral data, refer again to Table \ref{tab:spec_data}. 
Our primary molecular lines, which are used as a proxy for H$_2$, are the CO isotopologues in ALMA Band 6.
The $^{12}$CO, $^{13}$CO, and C$^{18}$O isotope observations capture gas kinematics and are used to quantify the source as well as trace the outflow.
As we observe the more rare isotopologues (i.e., from $^{12}$CO to $^{13}$CO to C$^{18}$O), the overall abundances and thus column densities drop. In many cases, this allows us to probe a larger fraction of the total column density of the structures if the lower column densities become optically thin in the rarer isotope. In effect, this means we may be able to see further down into the YSOs, through the envelope and into the disk. 

Given that one of the goals of eDisk is to characterize the properties of disks in young systems, the molecular line data were carefully chosen. C$^{18}$O is our main workhorse, as it traces the protostellar envelope and disk. $^{13}$CO also serves the same purpose as C$^{18}$O. $^{12}$CO traces outflows, and for our purposes is compared against disk tracers to avoid
outflow contamination to kinematic measurements of the disk and envelope. 
Although SO is typically detected in the shocked gas from bi-polar outflows, it is sometimes seen at the disk/envelope interface \citep{yen2014,sakai2014,ohashi2014}.

Visual inspection of our spectral cube data shows that we have significant emission in $^{12}$CO, $^{13}$CO, C$^{18}$O, and SO to warrant moment map analysis and position-velocity (PV) diagram analysis. All $c$-C$_3$H$_2$,  H$_2$CO, and DCN lines show faint $\gtrsim$3$\sigma$ emission, but the S/N is insufficient to trace the kinematics of the detected structures. Their channel and moment 0 maps are shown in the Appendix. And finally, SiO and CH$_3$OH only have upper-limits of  3$\sigma$.

Integrated intensity (moment 0) and intensity-weighted average velocity (moment 1) maps were created using the \texttt{bettermoments} Python package \citep{bettermoments}. 
Briefly, it integrates along the spectral dimension for each pixel and returns both a moment map and an error map. 
It is capable of determining the RMS from the empty channels in the cube and of determining the central velocity emission by fitting a quadratic around the brightest pixel and its 2 closest neighbors.
Threshold cuts were implemented to only include signal above 3$\sigma$.

In the following sub-sections, we discuss the general broad features of the main molecular line tracers ($^{12}$CO, $^{13}$CO, C$^{18}$O, and SO) with further analysis in \S \ref{gaskinematics}.

\subsubsection{\texorpdfstring{$^{12}$CO}{12CO} emission}
\begin{figure*}[h]
  \centering
    \includegraphics[width=1.0\textwidth]{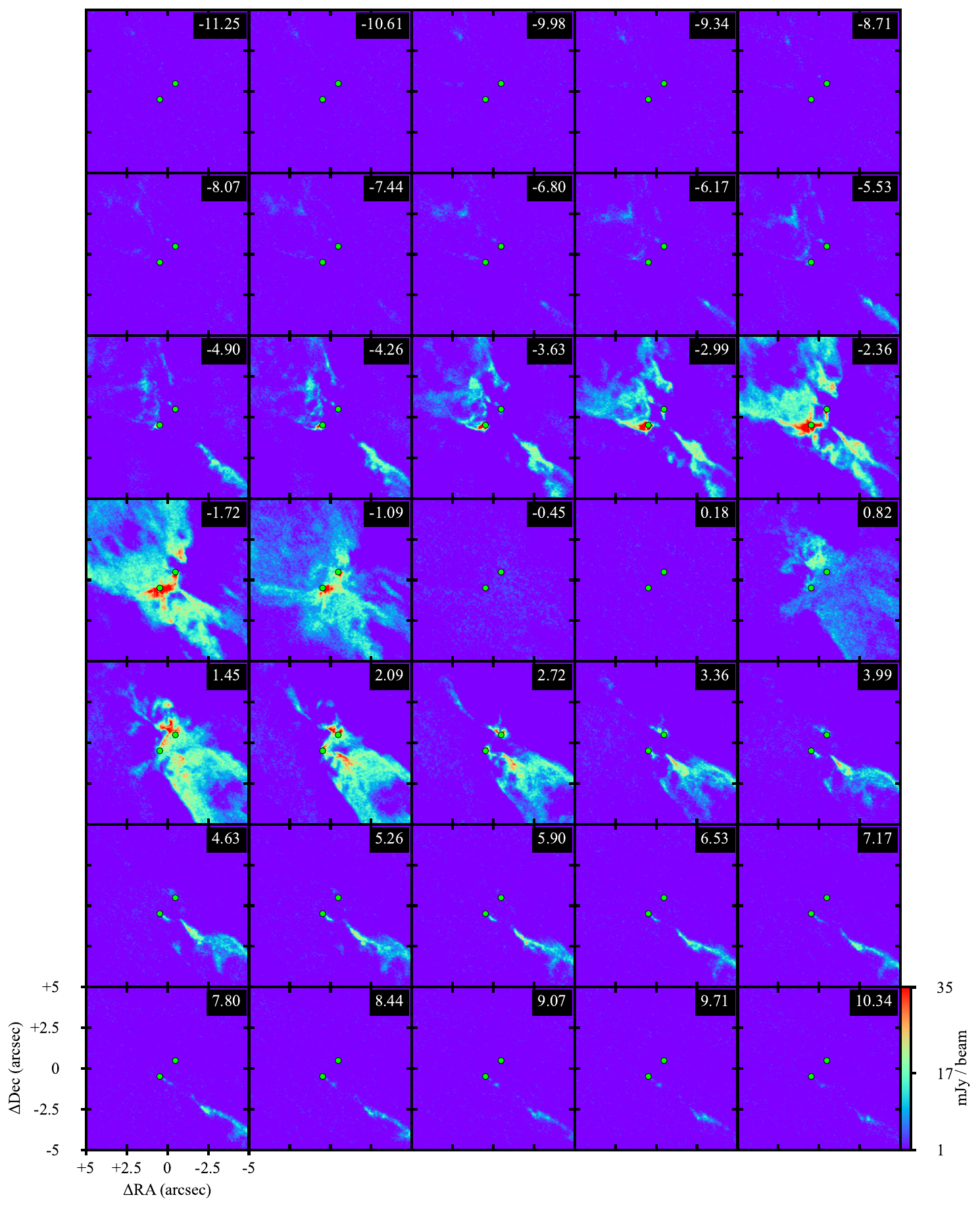}
  \caption{IRAS 32 $^{12}$CO (2-1) channel map with both short and long combined baselines and a Briggs robust parameter of 0.5. The velocity offset from 5.86 km s$^{-1}$ is shown in the top right of each channel. The minimum value of the colorbar is set to 1$\sigma$. Each individual image is within a 10 by 10 arcsecond box centered at RA=19:02:58.68 and DEC=-37:07:36.88. The beam size is 0$\farcs$11 $\times$ 0$\farcs$09 arcseconds with a PA of 87.2$^{\circ}$. The lime circular symbols represent the Gaussian-fit continuum peaks.}
  \label{fig:906c}
\end{figure*}

\begin{figure*}[h]
  \raggedleft
    \includegraphics[width=0.99\textwidth]{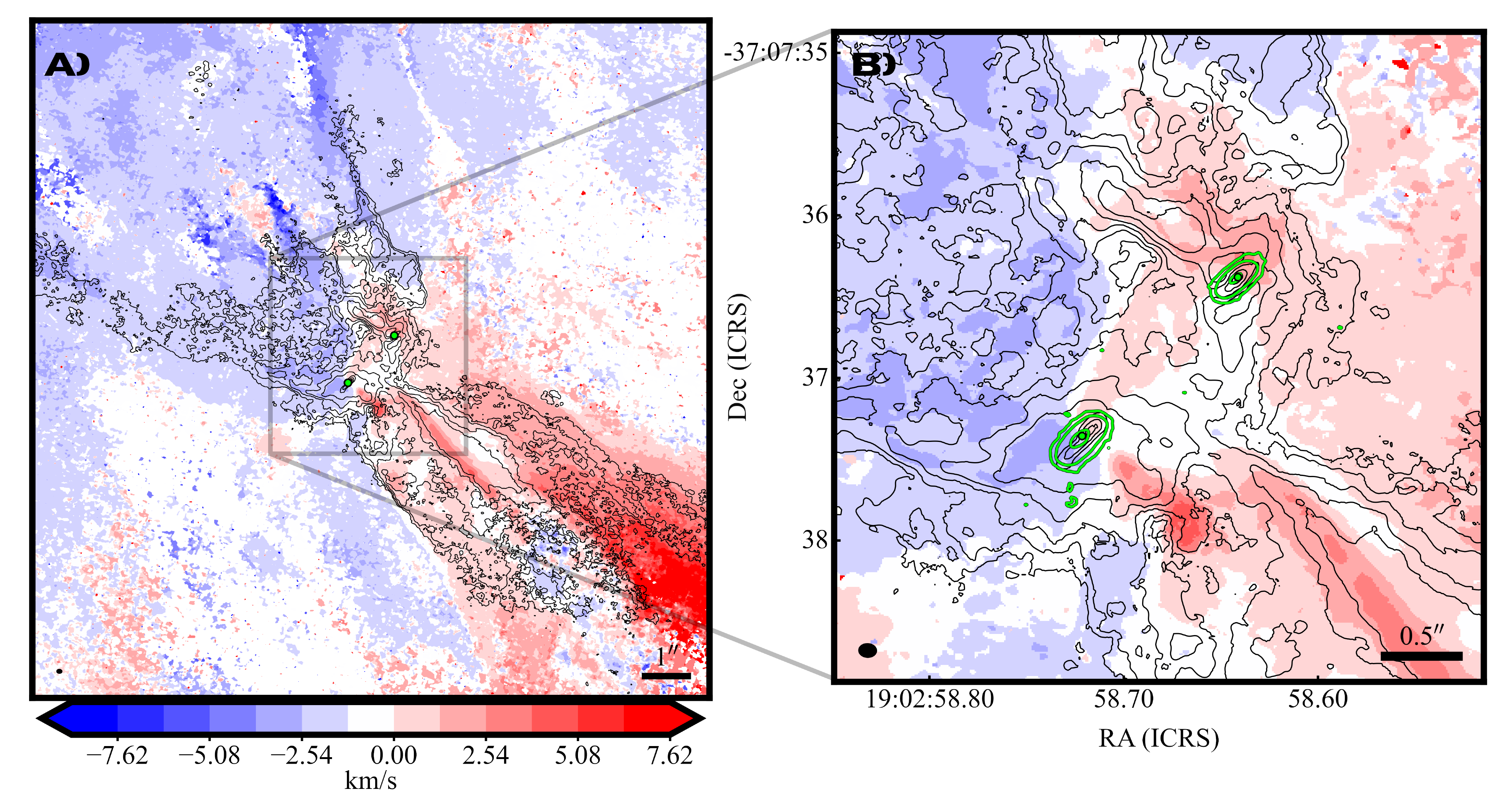}
    \includegraphics[width=0.99\textwidth]{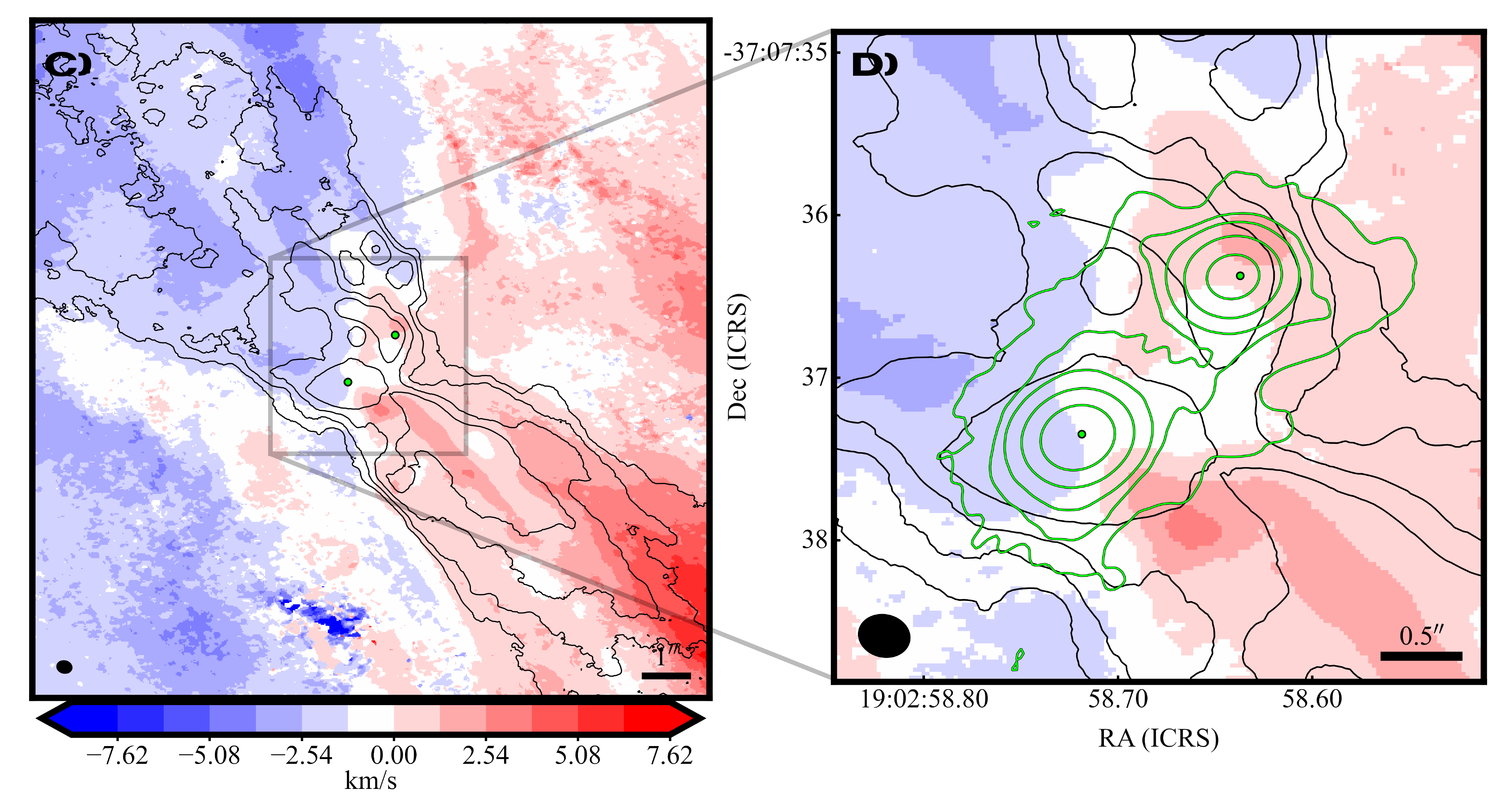}
  \caption{IRAS 32 $^{12}$CO Moment 1 map in the blue and red colormap with the adopted center velocity at 5.86 km s$^{-1}$. The black contours represent the Moment 0 data. The lime contours represent the 1.3 mm continuum data. The lime circular symbols represent the Gaussian-fit continuum peaks. The colormap color spacing is based on the $^{12}$CO channel spacing and offset by the systemic velocity of 5.86 km s$^{-1}$. The beam is in the bottom left corner in black, and a scalebar for comparison is in the bottom right corner. The top two images are of the combined long and short baseline data (with a Briggs robust parameter of 0.5) while the bottom two images are only the short baseline data (robust = 2.0). The left two images give an overview of the binary in a 14$^{\prime\prime}$x14$^{\prime\prime}$ window centered on the binary separation midpoint (from Table \ref{tab:cont_data}) while the right two images are in a 4$^{\prime\prime}$x4$^{\prime\prime}$ window.
See Table \ref{tab:spec_results} for details on the velocity ranges used to make the Moment 0 and Moment 1 maps.
{\bf Panel A:} The black contour levels are defined at 3$\sigma\sqrt{2^{n}}$ with $\sigma$=8.7 mJy km s$^{-1}$ and n$\in$[0,1,2,3,4]. The beam is 0$\farcs$11 $\times$ 0$\farcs$09 with a position angle of 87.2$^{\circ}$.
  {\bf Panel B:} The lime contour levels are defined at 3$\sigma$2$^{n}$ for $\sigma$=0.014 mJy and n$\in$[1,4,7].
  {\bf Panel C:} The black contour levels are defined at 3$\sigma\sqrt{2^{n}}$ with $\sigma$=67.4 mJy km s$^{-1}$ and n$\in$[0,1,2,3]. The beam is 0$\farcs$32 $\times$ 0$\farcs$26 with a position angle of 77.7$^{\circ}$.
  {\bf Panel D:} The lime contour levels are defined at 3$\sigma$2$^{n}$ for $\sigma$=0.076 mJy and n$\in$[1,3,5,9,13]. }
  \label{fig:901}
\end{figure*}

The CO emission toward IRAS 32 mainly traces the outflow and some circumstellar material.
Fig. \ref{fig:906c} shows the channel maps, and Fig. \ref{fig:901} shows the moment 1 maps for the combined observations. The systemic velocity adopted was 5.86 km s$^{-1}$, and the  velocities are all given with respect to this center velocity. 
The adopted systemic velocity was based on a visual inspection of the cube and will be used for all molecules hereafter, although the value will be revisited when analysing the PV diagrams in \S \ref{gaskinematics}.
The two circles mark the binary sources from Fig. \ref{fig:900} whereas the continuum emission is indicated by lime contours in Fig. \ref{fig:901}B and \ref{fig:901}D. The channel maps show structure that is consistent with two outflows (one from each source), seen especially in the blue emission toward the North. To the North there is also a possible outflow interaction region just to the northeast corner of the box in Fig. \ref{fig:901}A that demonstrates a wide range of velocities (blue and red). To the South, there is a clear structure that appears connected to the source B binary outflow component that is toward the blue side compared to the systemic velocity but then crosses over to the red.  Similarly, the A binary source has a redshifted component of its outflow that extends far into the red channels.  This may indicate that the two sources have slightly offset outflow ranges that smooths the emission from the individual outflows in Fig. \ref{fig:901}A. Nonetheless, in both the full dataset image Fig. \ref{fig:901}A and \ref{fig:901}B and the short baselines only Fig. \ref{fig:901}C and \ref{fig:901}D, one can still clearly see a velocity gradient along the individual disks and along the circumbinary structure that will be further examined below. 

\subsubsection{\texorpdfstring{$^{13}$CO}{13CO} emission}

\begin{figure*}[h]
  \raggedleft
    \includegraphics[width=0.99\textwidth]{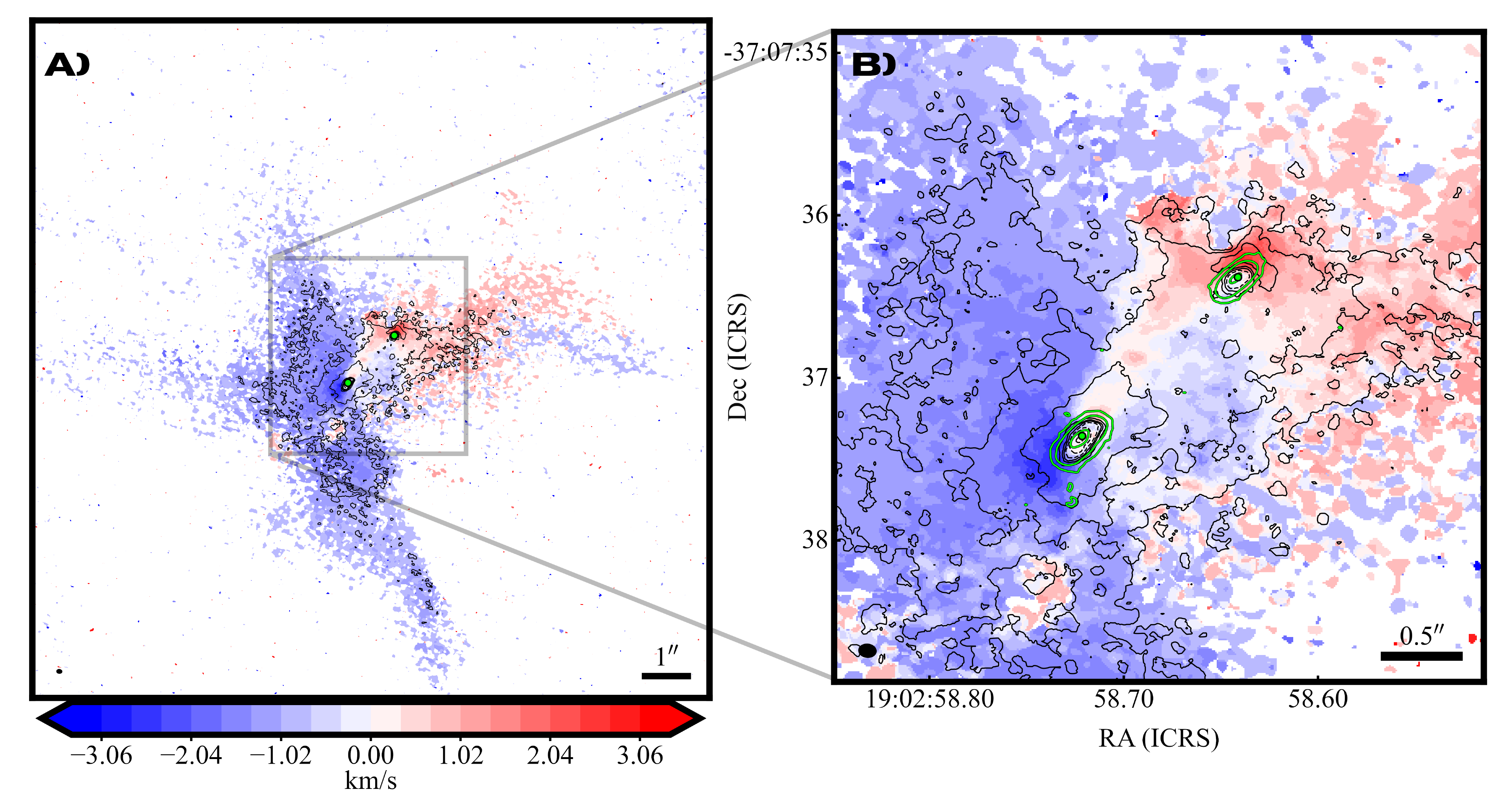}
    \includegraphics[width=0.99\textwidth]{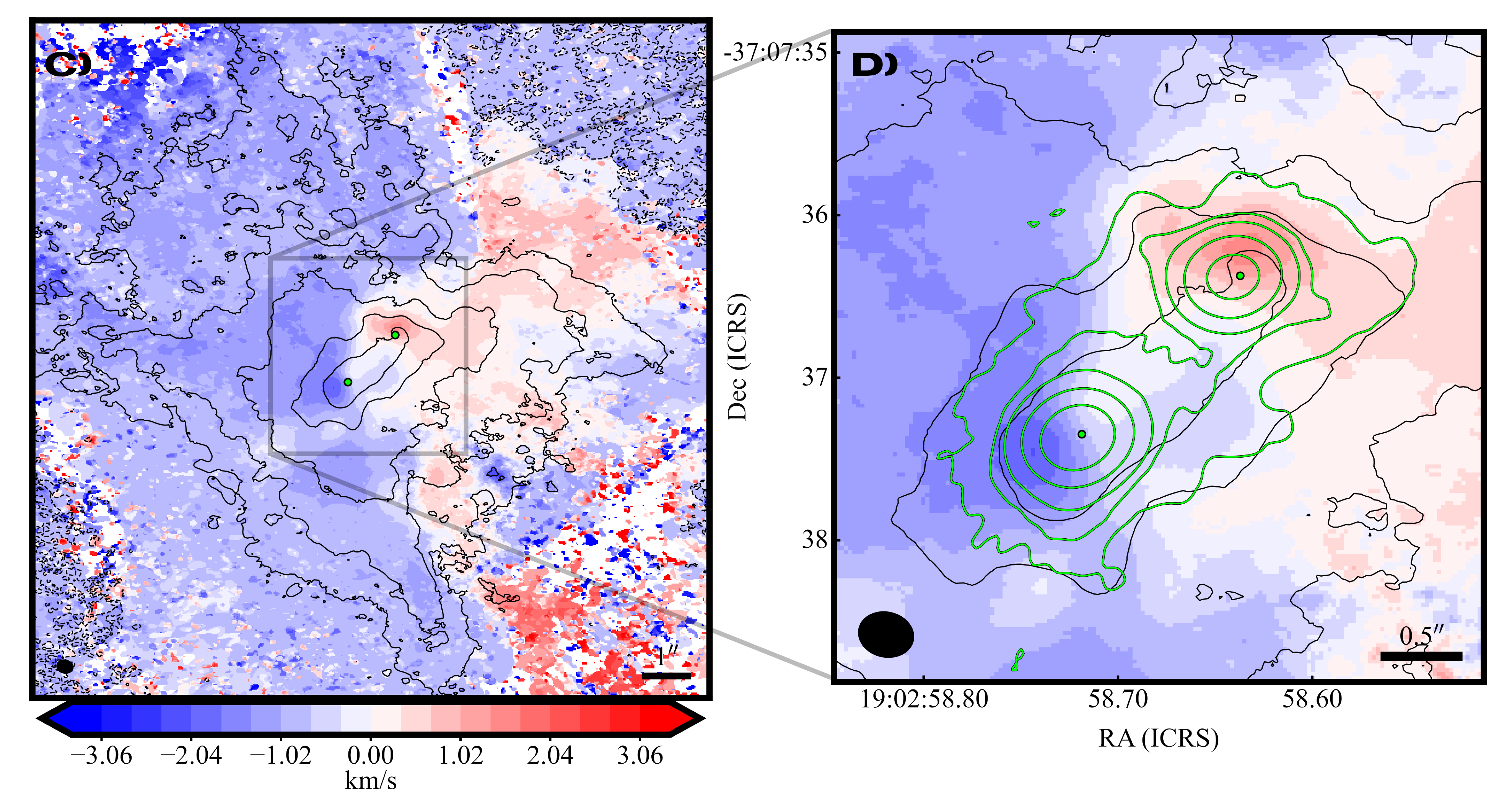}
  \caption{Same as Fig. \ref{fig:901} but for $^{13}$CO (2-1). 
  {\bf Panels A and B:} 
  The black dashed contour levels are defined at -3$\sigma$n with n$\in$[2,8,14] while the black solid contour levels are defined at 3$\sigma\sqrt{2^{n}}$ with $\sigma$=0.84 mJy km s$^{-1}$ and n$\in$[3,6,8,10,12,14]. The lime contours levels are defined at 3$\sigma$2$^{n}$ for $\sigma$=0.014 mJy and n$\in$[1,4,7]. The beam is 0$\farcs$11 $\times$ 0$\farcs$09 with a position angle of 84.2$^{\circ}$. 
  {\bf Panels C and D:} 
  The black dashed contour levels are defined at -3$\sigma$n with n$\in$[1,3] while the black solid contour levels are defined at 3$\sigma\sqrt{2^{n}}$ with $\sigma$=1.21 mJy km s$^{-1}$ and n$\in$[5,7,9,11,13]. The lime contours levels are defined at 3$\sigma$2$^{n}$ for $\sigma$=0.076 mJy and n$\in$[1,3,5,9,13]. The beam is 0$\farcs$34 $\times$ 0$\farcs$28 with a position angle of 77.6$^{\circ}$.}
  \label{fig:902}
\end{figure*}

Our moment map of $^{13}$CO, a molecule useful in tracing protostellar disks and the inner envelope, is shown in Fig. \ref{fig:902}. 
As seen in Fig. \ref{fig:902}A, the large scale emission appears to be tracing the higher density material around the outflow cavities of the combined two outflows from the binary, as well as the disks themselves.  There is an asymmetry in the southeast that may indicate asymmetry in the gas structure into which the outflow is impacting or streamer motion.  On the small scale, the inner regions of the disks are clearly seen in absorption in Figure \ref{fig:902}B as dashed contours (and later in the PV diagrams).  These negatives arise from the line center where the line is optically thick in contrast to the line wings that are optically thin in emission, all of which are common in young sources \citep[e.g.,][]{vantHoff2023,lin2023}.  
This absorption is likely from an over subtraction of the continuum due to the optically thick line emission (i.e. the compact continuum is not fully detected at line center so it is over subtracted), probably in combination with absorption by resolved-out, cold outer envelope material.
In the case of the eDisk source R CrA IRS 7B-a, which has an 
observed continuum peak brightness temperature of $\sim$195 K, a detailed 
radiative transfer modeling of the disk \citep{takakuwa2024} shows that the continuum over-subtraction is dominated by the high background dust brightness temperature and high dust optical depth.
On the other hand, there are also detected velocity gradients across each disk as well as a larger gradient over the circumbinary disk structure.

\subsubsection{\texorpdfstring{C$^{18}$O}{C18O} emission}

\begin{figure*}[h]
  \raggedleft
    \includegraphics[width=0.99\textwidth]{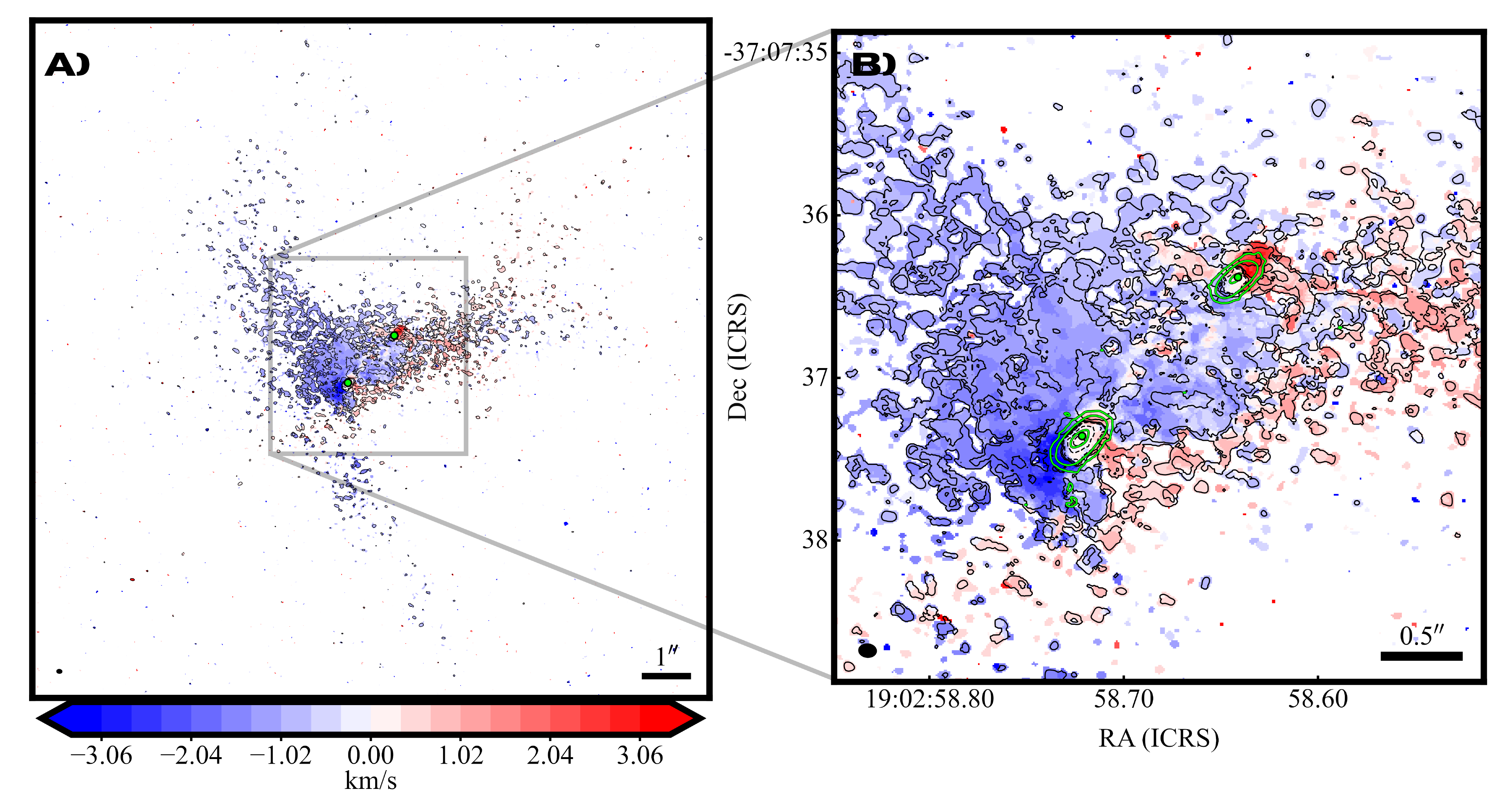}
    \includegraphics[width=0.99\textwidth]{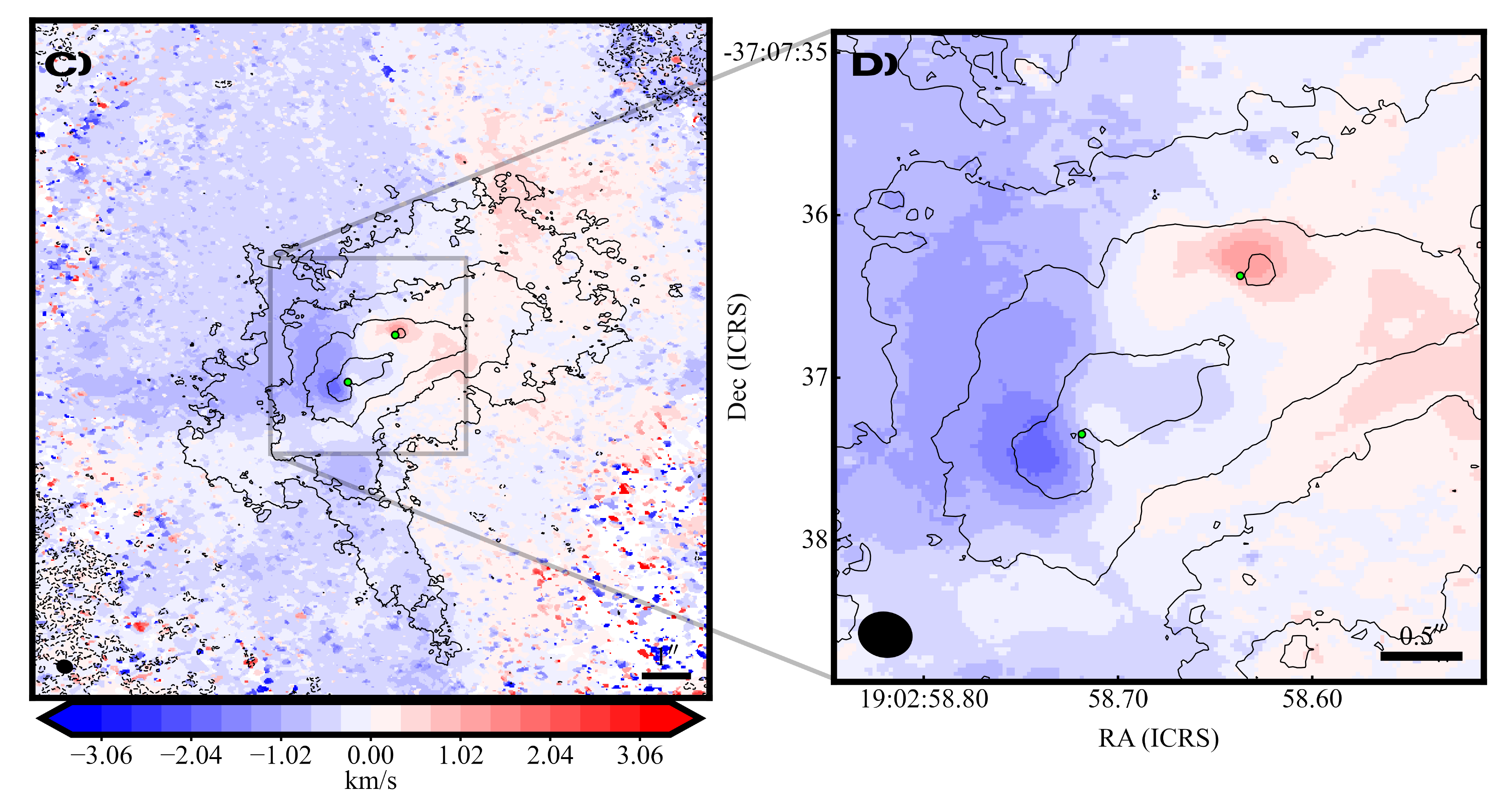}
  \caption{Same as Fig. \ref{fig:901} but for C$^{18}$O (2-1). 
  {\bf Panels A and B:} 
  The black dashed contour levels are defined at -3$\sigma$n with n$\in$[1,7,13] while the black solid contour levels are defined at 3$\sigma\sqrt{2^{n}}$ with $\sigma$=0.45 mJy km s$^{-1}$ and n$\in$[0,3,6,9,12]. The lime contours levels are defined at 3$\sigma$2$^{n}$ for $\sigma$=0.014 mJy and n$\in$[1,4,7]. The beam is 0$\farcs$11 $\times$ 0$\farcs$09 with a position angle of 84.9$^{\circ}$. 
  {\bf Panels C and D:} 
  The black dashed contour levels are defined at -3$\sigma$n with n$\in$[1,3] while the black solid contour levels are defined at 3$\sigma\sqrt{2^{n}}$ with $\sigma$=1.01 mJy km s$^{-1}$ and n$\in$[5,7,9,11,13]. The lime contours levels are defined at 3$\sigma$2$^{n}$ for $\sigma$=0.076 mJy and n$\in$[1,3,5,9,13]. The beam is 0$\farcs$33 $\times$ 0$\farcs$28 with a position angle of 77.8$^{\circ}$.}
  \label{fig:903}
\end{figure*}

In Fig. \ref{fig:903}, we show the C$^{18}$O moment 1 map overlaid with moment 0 and continuum contours. C$^{18}$O was one of the main targeted molecular lines in the eDisk program due to its ability to trace the protostellar disk from which we estimate disk rotation and morphology. 
These maps are very similar to the $^{13}$CO maps in Fig. \ref{fig:902}, although with less signal to noise.  
Again the C$^{18}$O is tracing the higher density material of the disks, the circumbinary structure, and around the outflow cavities of the combined two outflows. The blueshifted material to the southeast of the binary, particularly noticeable in Fig. \ref{fig:903}C, is spatially offset to east of the outflow redshifted material seen CO, Fig. \ref{fig:906c}.  On the disk scale, there is again some continuum over-subtraction.

\subsubsection{SO emission}
\label{SOemission}
\begin{figure*}[ht]
  \raggedleft
    \includegraphics[width=0.99\textwidth]{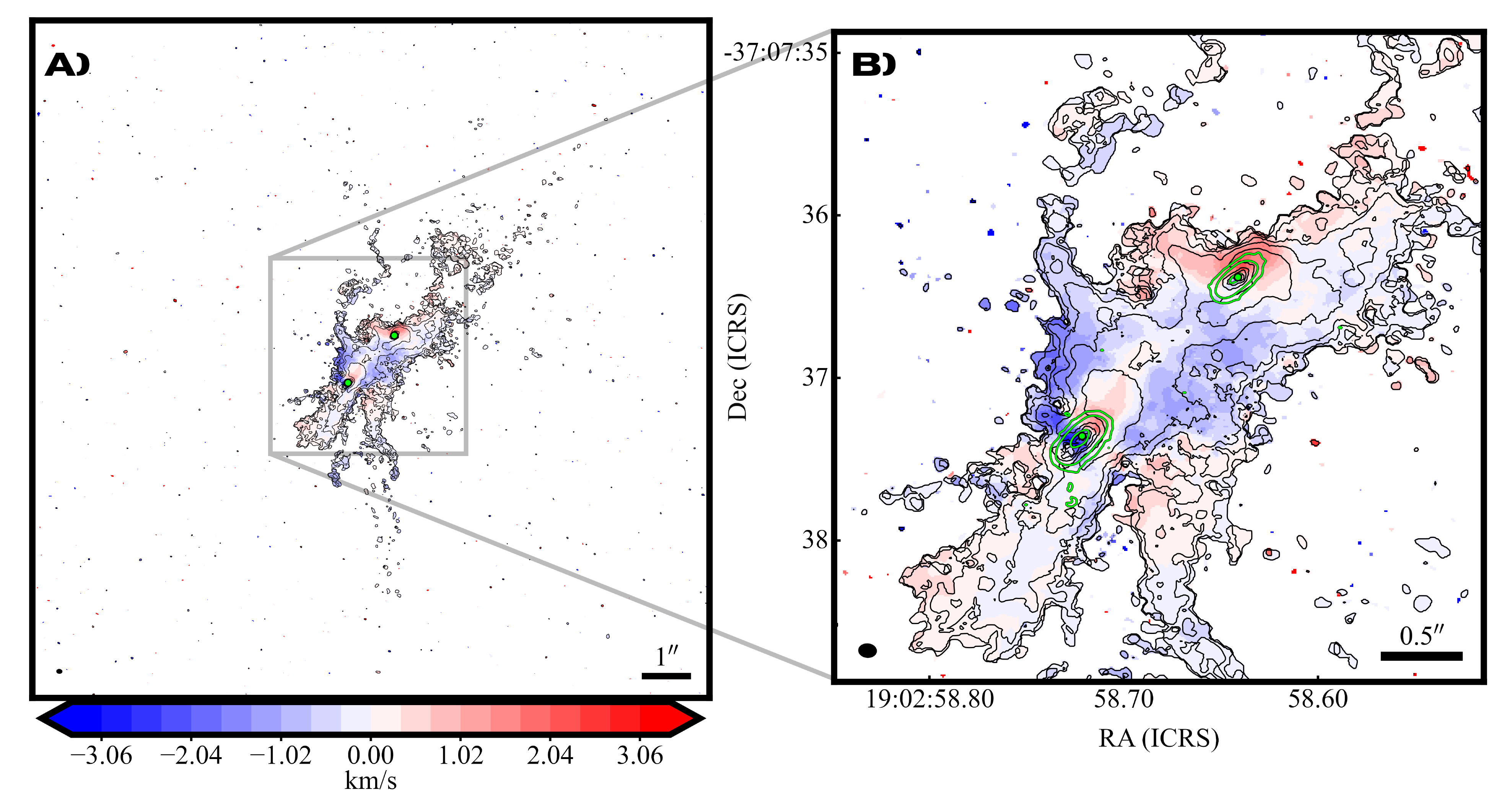}
  \caption{Same as Fig. \ref{fig:901} but for SO (6$_{5}$-5$_{4}$). 
The black solid contour levels are defined at 3$\sigma\sqrt{2^{n}}$ with $\sigma$=0.48 mJy km s$^{-1}$ and n$\in$[0,2,4,6,8,10]. The lime contours levels are defined at 3$\sigma$2$^{n}$ for $\sigma$=0.014 mJy and n$\in$[1,4,7]. The beam is 0.11$^{\prime\prime}$ x 0.09$^{\prime\prime}$ with a position angle of 85.1$^{\circ}$}
  \label{fig:904}
\end{figure*}

Similar to the previous three molecular lines, Fig. \ref{fig:904} shows the moment 0 and 1 maps of SO. 
The SO emission is morphologically and kinematically distinct compared to the CO isotopes in that it mostly traces the circumbinary structure and the circumstellar disks (and the inner envelope) with very little outflow emission. As can be seen in Fig. \ref{fig:904}B, the disks in SO have similar velocities with the exception of a strong blue-shifted component near the center of disk A. 

\begin{figure}[h]
  \centering
    \includegraphics[width=0.49\textwidth,trim=2 2 3 2,clip]{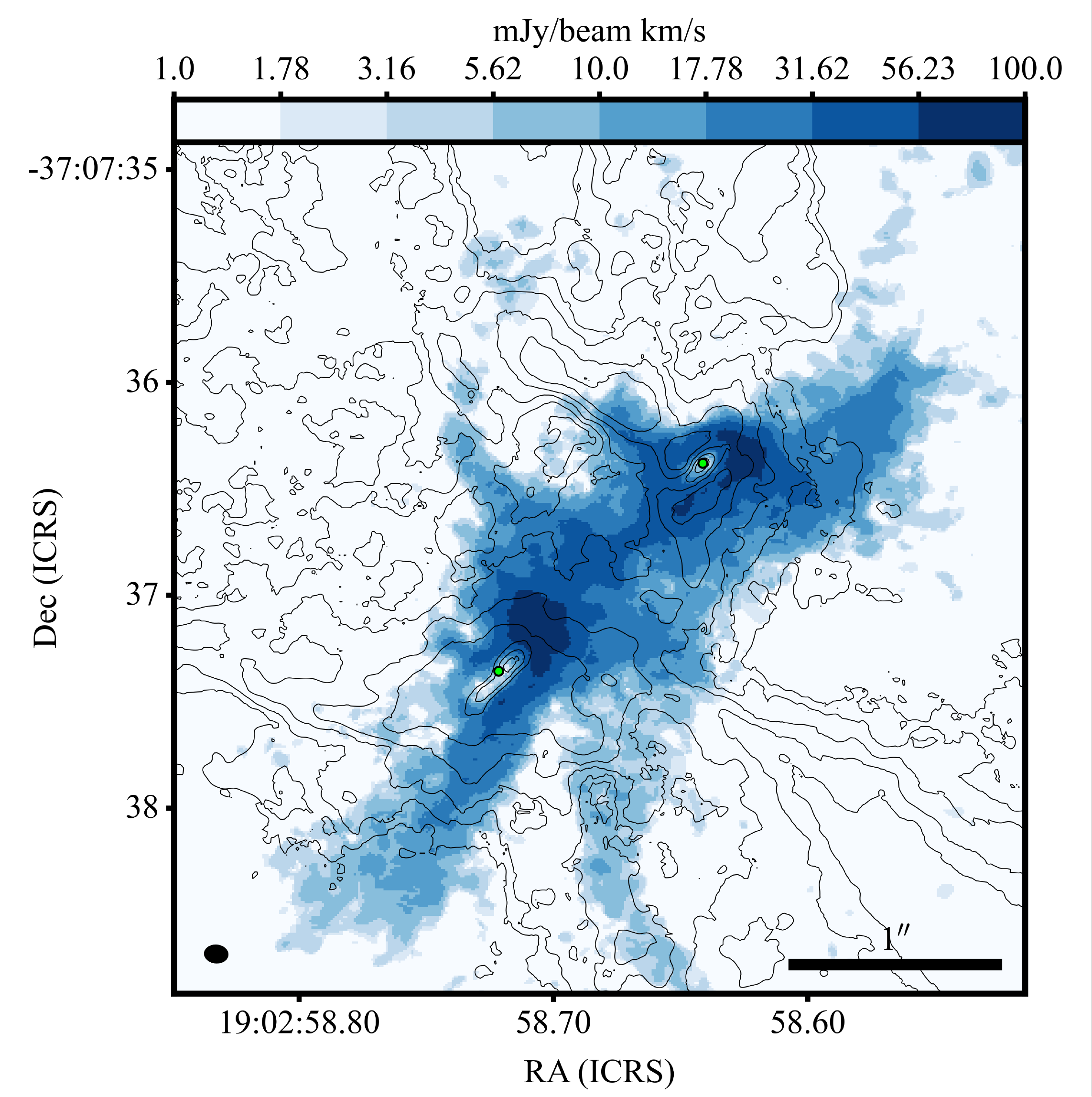}
  \caption{Overlay of $^{12}$CO moment 0 contours from Fig. \ref{fig:901}A and \ref{fig:901}B with the SO moment 0 from Fig. \ref{fig:904}B but shown as a colormap that emphasizes the filamentary structures. 
  }
  \label{fig:904b}
\end{figure}

With Fig. \ref{fig:904b}, we can compare the $^{12}$CO moment 0 from Figs. \ref{fig:901}A and \ref{fig:901}B with the SO moment 0 from Fig. \ref{fig:904}.
In general, the SO along the midplane appears to be tracing the circumbinary disk and the circumstellar disks. SO is also tracing some of the jet or outflow material very close to the protostars, particularly seen to the northeast along the outflow direction of source B in Figs. \ref{fig:904b} and \ref{fig:904}B. Note that this emission is also seen in $^{13}$CO in Fig. \ref{fig:902}B, and in both cases it is redshifted.  In addition, there are two SO  filamentary structures to the northeast and southwest in Fig. \ref{fig:904b} that appear to begin between the protostars and curve away. The southwest SO filament is aligned with a depression in the $^{12}$CO moment 0 contours, which is slightly offset to the west of the $^{13}$CO traced cavity wall seen in Figs. \ref{fig:902}A and B.
Although the two filamentary structures are not tracing the same material as the $^{12}$CO directly, they may be still related to the outflow structures (e.g., enhanced cavity walls or interfaces) or perhaps streamers shaped by the outflow.

\subsubsection{Other molecules}

We have also imaged the other molecular lines seen in Table \ref{tab:spec_data}. We detected $>$3$\sigma$ emission in all cases except SiO, $c$-C$_3$H$_2$ (5$_{2,4}$-4$_{1,3}$), and CH$_{3}$OH, where we instead only report their 1$\sigma$ noise levels (0.63 and 0.52 mJy beam$^{-1}$, respectively). See the details of the images listed in Table \ref{tab:spec_results}. The channel maps of the detected lines not shown in the main text are provided in the Appendix.

\begin{table*}[h]
    \centering
    \caption{Spectral Analysis Overview} \label{tab:spec_results}
    \begin{tabular}{l c c c c c c c}
\hline \hline
Molecule            & Maps          & \multicolumn{2}{c}{RMS$^a$}               & \multicolumn{2}{c}{Moment 0$^b$} &  \multicolumn{2}{c}{Moment 1$^c$}\\ 
(GHz)               &                       & \multicolumn{2}{c}{(mJy beam$^{-1}$)} & \multicolumn{4}{c}{Velocity Range (km s$^{-1}$)}  \\
\hline
SiO                 & Not detected          & 0.63   & --    & --              & --              & --               & --             \\
DCN                 & Channel \& Moment     & 0.64   & --    & -2.46 - +1.56   & --              & --               & --             \\
$c$-C$_3$H$_2$ 217.82 & Channel \& Moment   & 0.60   & --    & -2.46 - +1.56   & --              & --               & --             \\
$c$-C$_3$H$_2$ 217.94 & Channel \& Moment   & 0.54   & --    & -2.46 - +1.56   & --              & --               & --             \\
$c$-C$_3$H$_2$ 218.16 & Not detected        & 0.55   & --    & --              & --              & --               & --             \\
H$_{2}$CO 218.22    & Channel \& Moment     & 0.53   & --    & -3.80 - +2.90   & --              & --               & --             \\
CH$_3$OH            & Not detected          & 0.52   & --    & --              & --              & --               & --             \\
H$_{2}$CO 218.48    & Channel \& Moment     & 0.53   & --    & -2.46 - +2.90   & --              & --               & --             \\
H$_{2}$CO 218.76    & Channel \& Moment     & 1.59   & --    & -2.50 - +1.51   & --              & --               & --             \\
C$^{18}$O           & Channel \& Moment     & 1.68	 & 3.1   & -5.01 - +4.50   & -5.35 - +4.67   & -5.85 - +5.34    & -6.18 - +5.51  \\
SO                  & Channel, Moment \& PV & 2.06	 & --    & -4.35 - +4.34   & --              & -5.18 - +5.17    & --             \\
$^{13}$CO           & Channel, Moment \& PV & 2.20   & 4.1   & -5.68 - +5.01   & -5.85 - +5.51   & -6.52 - +5.84    & -6.68 - +6.34  \\
$^{12}$CO           & Channel, Moment       & 1.00   & 1.9   & -18.23 - +20.50 & -31.57 - +30.03 & -21.45 - +23.68  & -8.08 - +8.00  \\

\hline
\end{tabular}
\tablenotetext{}{}
    \tablenotetext{}{$^a$ RMS values for the channel maps (short and long baselines combined and short baseline only). $^b$ Range used to make the moment 0 map with a 4$\times$RMS cutoff (short and long baselines combined and short baseline only). $^c$ Range used by \texttt{bettermoments} to measure the moment 1 velocities (short and long baselines combined and short baseline only).} 
\end{table*}

\section{Analysis and Discussion} \label{sec:discussion}

\subsection{Continuum Emission} \label{cont-asym}

As is clear in Fig. \ref{fig:900}, the two circumstellar disks of the system are well aligned parallel to each other.  In addition, the two disks are also well aligned with the (projected) orbital plane of the circumbinary structure, as seen in Figs. \ref{fig:900} (left) and \ref{fig:901}B.
This suggests that
this particular binary system likely formed in a relatively ordered process, such as disk fragmentation instead of a less ordered formation process, such as turbulent fragmentation followed by orbital migration to match the close spacing. 

In addition, as hinted in Fig. \ref{fig:900} and shown with line projections in Fig. \ref{fig:905c}, there are asymmetries in the 1.3~mm continuum disk emission. Specifically, these asymmetries are present in both disks and present themselves as more emission in 
the northeast half of the disk than in the southwest half; indeed, in both cases, the emission peak is shifted to northeast along the minor axis. 
We can use the outflows to determine the near/far side of the disks. Since the brighter (northeast) side of the disks are superposed on the blueshifted side of the CO outflow (see Fig. \ref{fig:901}), that is the far side of the disks. Thus, this brightness asymmetry along the minor axis is evidence that the dust disk is optically thick and has a significant geometrical thickness with the continuum-emitting dust not well-settled to the disk midplane, similar to other eDisk sources 
(e.g., IRAS 04302+2247, \citealt{lin2023}; R CrA IRS7B-a,  \citealt{takakuwa2024}).




\subsection{Gas Kinematics}\label{gaskinematics}

To analyze the gas kinematics in the binary system, we follow a similar procedure to \cite{edisk} using position-velocity (PV) diagrams. The PV diagrams were created using the CASA task \texttt{impv}, and the fits to the data were done using the \texttt{pvanalysis} package of the SLAM (Spectral Line Analysis/Modeling) Python library \citep{yusuke_aso_2023}.  

SLAM uses the gas kinematics to fit the gas rotational profile as a simple power-law with radius described as 

\begin{equation}
    v = V_b\left(\frac{r}{R_b}\right)^{-p},
\end{equation} 

\noindent where $V_b$ is the mean velocity, $R_b$, is the radius of the mean velocity, and $p$ is the power-law index.
A power-law index of $p\approx0.5$ would be consistent with Keplerian rotation of a disk, whereas  $p\approx1$ would be consistent with rotating and infalling material with conserved angular momentum.  One can estimate the Keplerian mass enclosed within $R_b$ by 
\begin{equation} 
    M = \frac{V_b^2 R_b}{G \sin^2 i},
\end{equation}

\noindent where $G$ is the gravitational constant and $i$ is the inclination angle from the continuum fitting. SLAM will fit the velocities along the emission edge and ridge of the contours (see Fig. \ref{fig:907c2}).  These two cases should bracket the Keplerian central protostellar mass, assuming a Keplerian profile power-law. For the angle of the PV cuts, we used the Gaussian-fit position angles. The PV line widths used are $\sim$1 beam for all images and any subsequent analysis. Note that IRAS 32, with the combination of two close disks and a large circumbinary disk/envelope, complicates the fitting process. In addition, the fitting process is also more complicated for young sources with evidence of infall from the envelope onto the circumstellar disks in PV diagrams \citep[see][and specifically their Fig. 10] {ohashi1997}. 

\subsubsection{Circumbinary structure}

\begin{figure}[h]
  \centering
    \includegraphics[width=0.45\textwidth,trim=0 2 2 2,clip]{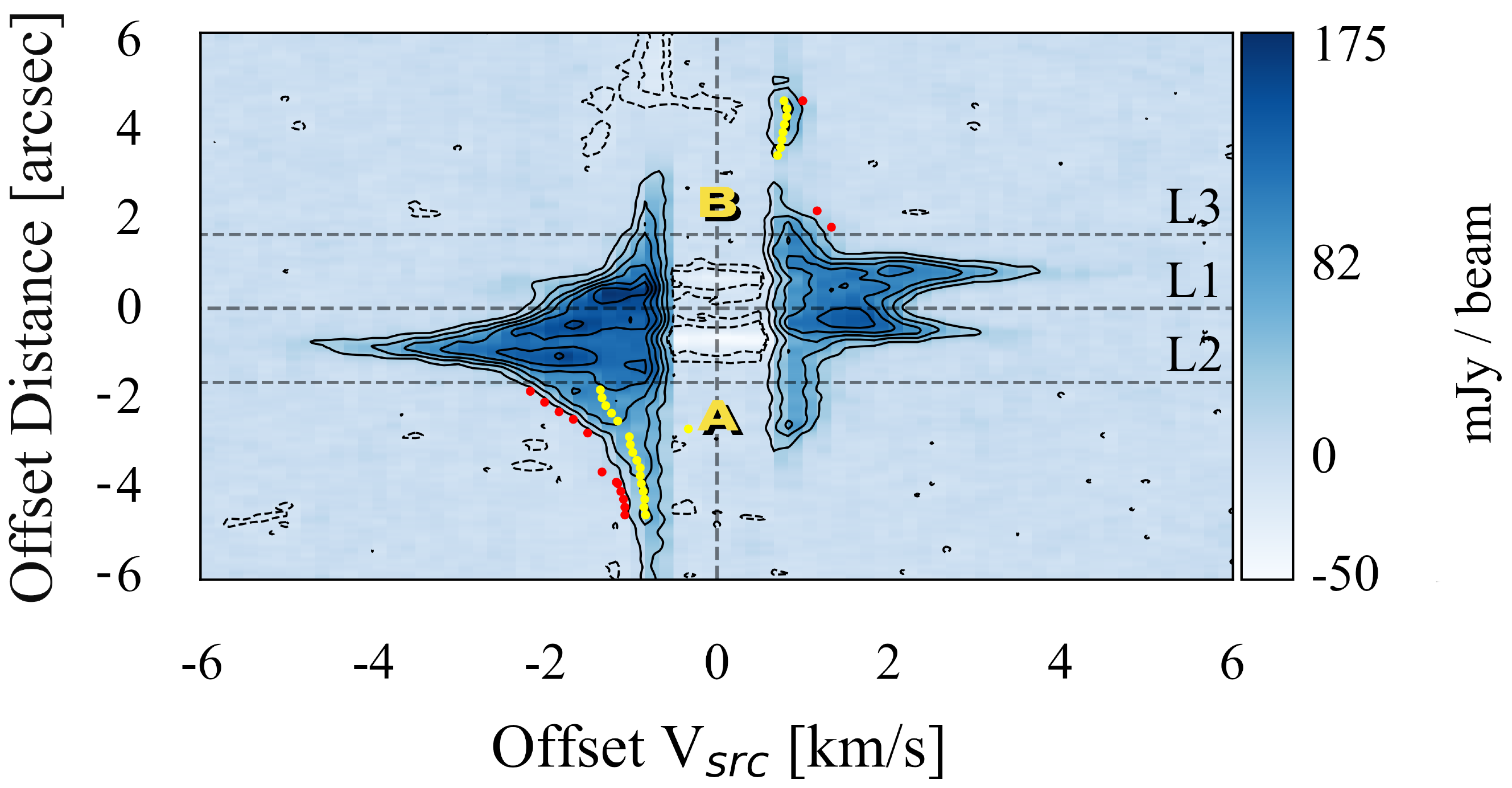}
  \caption{IRAS 32 Position-Velocity (PV) diagram of the $^{13}$CO molecular line in short-baseline only. The dots represent the SLAM ridge fit in yellow and the SLAM edge fit in red. The PV cut was taken along the binary separation from A to B, passing through the Gaussian-fit centers. The thickness of the cut was approximately 1 beam major axis. The black contours are at 3$\sigma$n for $\sigma$=3 mJy beam$^{-1}$ and n$\in$[-2,-1,3,6,9,12,15,18]. The separation between the binary sources is 1.38$^{\prime\prime}$. The adopted center velocity is 5.86 km s$^{-1}$. L1, L2, and L3 label the Lagrange points, assuming equal masses for the binaries. 
  }
  \label{fig:907c2}
\end{figure}

For the circumbinary structure seen in Figs. \ref{fig:901}D and \ref{fig:902}D, the $^{13}$CO emission provides the best compromise between signal to noise and outflow contamination.  
Fig. \ref{fig:907c2} shows the $^{13}$CO PV diagram for the circumbinary structure using the short baseline observations only. Although one can clearly see evidence of differential rotation, there is also emission in the off-Keplerian quadrants (upper left and bottom right) of the diagram, which suggests infall motions too.  In addition, those quadrants also have the peaks from the circumstellar disks, but we only fit the larger velocity variations for the outer region of the circumbinary structure. 
Table \ref{tab:slam_params} gives the SLAM output fits for Fig. \ref{fig:907c2}.  Note that we allowed the central velocity to be a fit parameter in SLAM and did not force the adopted 5.86 km s$^{-1}$ used in our figures.
The fit values of p were 0.58  and 0.79 for the ridge and edge fits, respectively.  This is suggestive of some combination of Keplerian and infall motions.  Although not conclusive, we posit that the circumbinary structure is a circumbinary disk. This may be fed by the filamentary structures seen in SO, Fig. \ref{fig:904}.
Circumbinary disks surrounding two circumstellar disks are frequently seen in young systems \citep[e.g.,][]{looney1997,takakuwa2012,hsieh2020,maureira2020,diaz-rodriguez2022}. Although the SLAM fits do not provide strong evidence of the Keplerian kinematics nor place very accurate mass limits, the general circumbinary shape and kinematics follow the trends seen in other young sources. 

\begin{center}
\begin{table}
    \centering
    \caption{SLAM Fit Parameters} \label{tab:slam_params}
    \begin{tabular}{l l l l} 
\hline \hline
 $^{13}CO$              & Circumbinary      & Source A          & Source B    \\
\hline
Ridge                   &                   &                   &                       \\
\hline
M$_b$ (M$_{\odot}$)     & 0.53$\pm$0.09     & 0.5$\pm$0.6     & 0.48$\pm$0.03         \\
v$_{sys}$ (km s$^{-1}$) & 5.809$\pm$0.004     & 5.05$\pm$0.01     & 4.87$\pm$0.01         \\
p                       & 0.58$\pm$0.02     & 0.35$\pm$0.01     & 0.4$\pm$0.2         \\
\hline
Edge                    &                   &                   &                       \\


\hline
M$_b$ (M$_{\odot}$)     & 1.0$\pm$0.9     & 0.91$\pm$0.01     & 0.32$\pm$0.01         \\
v$_{sys}$ (km s$^{-1}$) & 5.791$\pm$0.005     & 5.48$\pm$0.01     & 5.46$\pm$0.01         \\
p                       & 0.79$\pm$0.02     & 0.36$\pm$0.01     & 0.71$\pm$0.02         \\

\hline
\end{tabular}
\tablenotetext{}{}
\end{table}
\end{center}

\subsubsection{Circumstellar Disks}

\begin{figure}[h]
  \centering
    \includegraphics[width=0.49\textwidth,trim=2 2 2 2,clip]{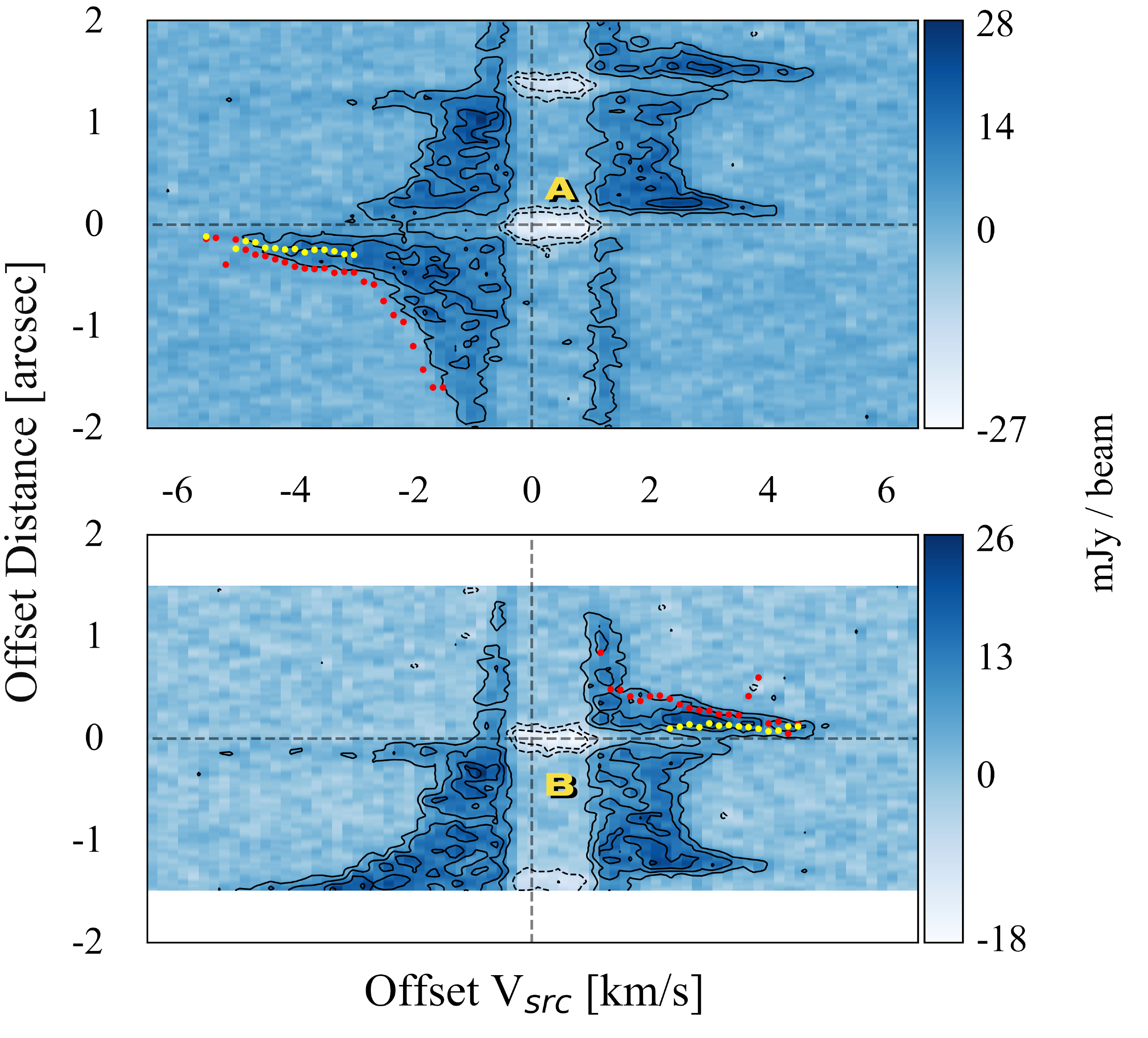}  
  \caption{Same as Fig. \ref{fig:907c2} except in the combined short- and long-baselines. The black contours are at 3$\sigma$n for $\sigma$=2 mJy beam$^{-1}$ and n$\in$[-2,-1,1,2,3,4]. Only the outer portion is fit due to larger uncertainties in the region between the sources. The window size was chosen to be identical in both panels. {\bf Top panel:} Centered on Source A. {\bf Bottom panel:} Centered on Source B.}
  \label{fig:907b}
\end{figure}

There are signs of circumstellar disk rotation present to some extent in all of our disk maps. 
As can seen in Fig. \ref{fig:907b}, there is absorption in the center of the disks at the low velocities of the line center, but that does not limit our ability to fit Keplerian velocity profiles as the higher velocities in the inner disk are more constraining. 
Instead, the two larger issues for fitting velocity profiles in the PV diagrams of Fig. \ref{fig:907b} are that the  binary systems kinematically interact (i.e., overlapping spatially and spectrally) and, more importantly, that there is contamination from circumbinary, envelope, and  outflow velocity components, especially in $^{12}$CO.  

On the other hand, the data for $^{13}$CO and C$^{18}$O, which also have absorption in the inner region, are more consistent with each other and trace slightly closer into the disk for both sources.  
Nonetheless, even in these cases, the effect of the other binary system is demonstrated in the gas kinematics.  For both sources, the inner portion of their PV diagram has a lower gas kinematic velocity that makes it more difficult to fit a protostellar mass with the kinematics. 

We chose to use the mass fit from the $^{13}$CO emission, e.g., Fig. \ref{fig:902}A and B as the best compromise between signal to noise and contamination from the outflow and other disk. To minimize the effect of the other disk, we only fit the opposite side of the disks, which limits our fitting constraints.  
The $^{13}$CO PV diagrams of A and B differ in the length of the PV cuts, due to considerations of contamination from the other binary source since the position angle of both sources also coincides with the separation angle of the sources. For those reasons, we fit the IRAS 32 A PV diagram extended to 2 arcseconds around the center, while the IRAS 32 B fit extended  out to 1.5 arcseconds around the center. This of course means some overlap in the inner regions, as the binary separation is 1.38 arcseconds. In addition, as we sample the gas to the outside of the disk, the kinematics will be more and more affected by the combined mass of the system, which will also change the rotation curve, making the kinematics more difficult to fit.

Table \ref{tab:slam_params} gives the SLAM output fits for Fig. \ref{fig:907b}.  The disk A fit values for p were 0.35 and 0.36 for the ridge and edge  fits, respectively.  However, the mass fits for source A are different (0.5 and 0.9 M$_\odot$, respectively) with very large uncertainty for the ridge fit. 
The disk B fit values for p were 0.43 and 0.71 for the ridge and edge  fits, respectively. This is closer to the expected values for  Keplerian motion, but as can be seen in Fig. \ref{fig:907b}, the ridge fit is likely not a good representation of the kinematics as it falls much more steeply than the edge fit. The mass for the ridge only fit is 0.48$\pm$0.03 M$_\odot$.



\subsection{Inner binary region}

In all 4 moment maps (Fig. \ref{fig:901}, \ref{fig:902}, \ref{fig:903}, \& \ref{fig:904}), the inner region along the binary separation is at or nearly at the system velocity with few velocity fluctuations. This is interesting because the PV diagrams show that 1) there is substantial infalling material all along the binary, assumed to be from either the large mass reservoir in the envelope or the circumbinary disk, and that 2) the material in the Keplerian disks are close enough that there could be kinematic interactions between them. 

Towards the first point, although infall is expected to favor the lower mass protostar in young binary systems around rotating gas reservoirs \citep[e.g.,][]{bate2000}, the similar protostar masses suggest infall all along the binary, including the inner region. Nevertheless, such a presumably chaotic region is not seen in $^{12}$CO, $^{13}$CO, nor C$^{18}$O at our current velocity and spatial resolution. The PV diagrams show an equal amount of material on both sides around the rest velocity in the region between the binaries, which is likely from the circumbinary disk with its own Keplerian rotation. This is not the case with SO in Fig. \ref{fig:904};  while there is a small central region at the systematic velocity, there is also a velocity gradient along the binary separation. In the PV diagram, material seems to favor the negative velocities, or the blue side. Although the SO emission profile generally matches the CO isotopologues that trace the circumstellar disk rotation, SO is commonly found to trace 
multiple features of a protostar, including the disk, warm inner envelope, low-velocity outflow,  jets, and binary interaction flows \citep[e.g.,][]{tychoniec2021,takakuwa2020}.

\begin{figure}
  \centering
    \includegraphics[width=0.47\textwidth]{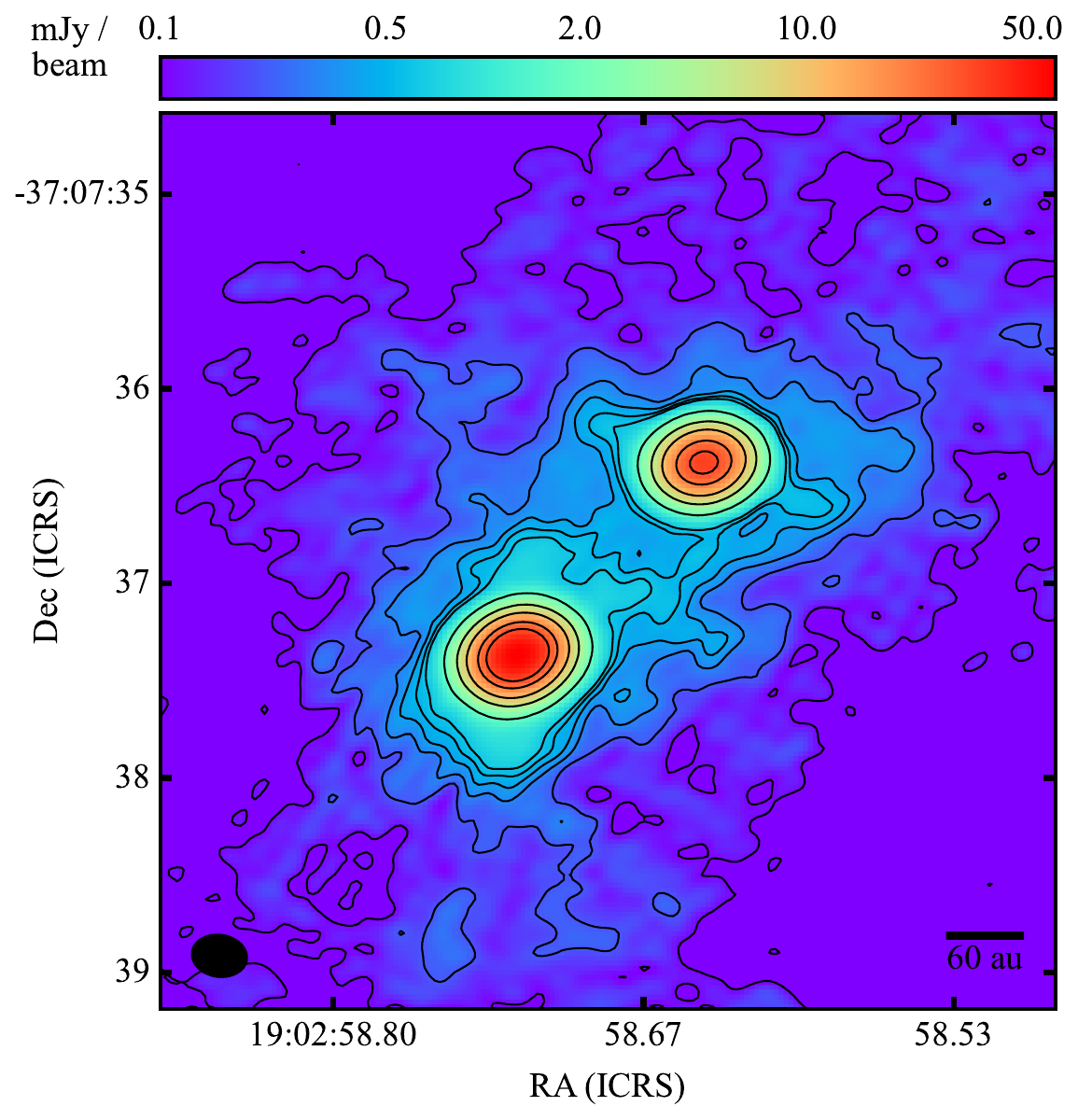}
  \caption{Same as Fig. \ref{fig:900}, short-baseline but with a Briggs robust parameter of 1.0. Contour levels are shown at 3$\sigma n$ for $\sigma$=0.0338 mJy beam$^{-1}$ and n$\in$[1,2,3,4,5,6,18,50,100,200,300]. The beam has a major and minor axis of 0$\farcs$286 $\times$ 0$\farcs$217, respectively, with a PA of 81.7$^{\circ}$.}
  \label{fig:900b}
\end{figure}


Finally, using only the short-baseline continuum data with a robust parameter of 1.0, shown in Fig. \ref{fig:900b}, we detect continuum emission between the two circumstellar disks. Additionally, the strong SO emission in Fig. \ref{fig:904b} between the binaries, which is also seen in the moment 0 map in Fig. \ref{fig:904}B, also shows the connection of gas between the binaries.
This emission could be evidence of gas funneling between the disks, as previously seen in L1551 IRS 5 \citep{takakuwa2017,takakuwa2020} and predicted by numerical simulations \citep[e\.g.,][]{matsumoto2019}.

\subsection{Outflows}

Among the four molecular lines chosen for analysis, $^{12}$CO, to a lesser extent $^{13}$CO, and arguably even C$^{18}$O, capture emission from the outflows (see Fig. \ref{fig:901} \& \ref{fig:902}). That is, we see high-velocity emission normal to the disk PA for disk A, disk B, and the circumbinary disk. The driving mechanism for the overall outflows in this system could be all 3 disk components. Indeed, MHD simulations of \cite{machida2009} revealed that  circumbinary disks of a protostellar system can drive outflows, although they examined a closer binary separation.

While the $^{12}$CO moment maps show clear evidence of an outflow originating from disk A, the case for disk B is best seen in Fig. \ref{fig:906c} in the blue channels between the velocity ranges of -4.90 and -2.99 km s$^{-1}$, and the red channels between the velocity ranges of +1.45 and +2.72 km s$^{-1}$.
However, in Figs. \ref{fig:901}A there is also clear redhifted material in the northern blueshifted lobe and blueshifted material in the southern redshifted lobe.

For the possibility of the circumbinary structure driving an outflow, it is less clear.  
As seen in Fig. \ref{fig:906c} in the northeast blue channels (velocity ranges of -6.17 to -5.53 km s$^{-1}$) and in the southwest (velocity ranges of +3.36 to +9.71 km s$^{-1}$), there are strong persistent features that are equidistant to the two disks. These may be interaction regions (between the A and B outflows, see below), so visually the existence of three outflows is inconclusive. On the other hand, there is also widespread low-velocity outflow gas emission that is best seen in the blue channels in the north traced by CO and $^{13}$CO,  emphasized in Figs. \ref{fig:901}A and \ref{fig:902}A. In either case, further modeling is necessary to address a third outflow.

For the clear individual protostellar outflows, the blue-shifted (negative velocities relative to the system velocity) regions tend to be in the northeast for both the individual sources and the circumbinary disk. Similarly, the red-shifted regions are opposite, in the southwest. They show both low- and high- velocity emission centered on both sources and centered on the inferred circumbinary disk center. The outflow 3$\sigma$ emission extends to the edges of the $^{12}$CO field-of-view, putting the projected length of any one particular side at about 20 arcsec (3000 au). We calculate a deprojected dynamical age of the observed outflow of $\sim$280 years by using the deprojected maximum velocity of the $^{12}$CO outflow (adopting an inclination of 70$^\circ$, the average of the disk inclination fits from Table \ref{tab:cont_derived_data} and $\sim$17 km s$^{-1}$ from visual inspection of the cube).

\begin{figure}[h]
  \centering
    \includegraphics[width=0.45\textwidth]{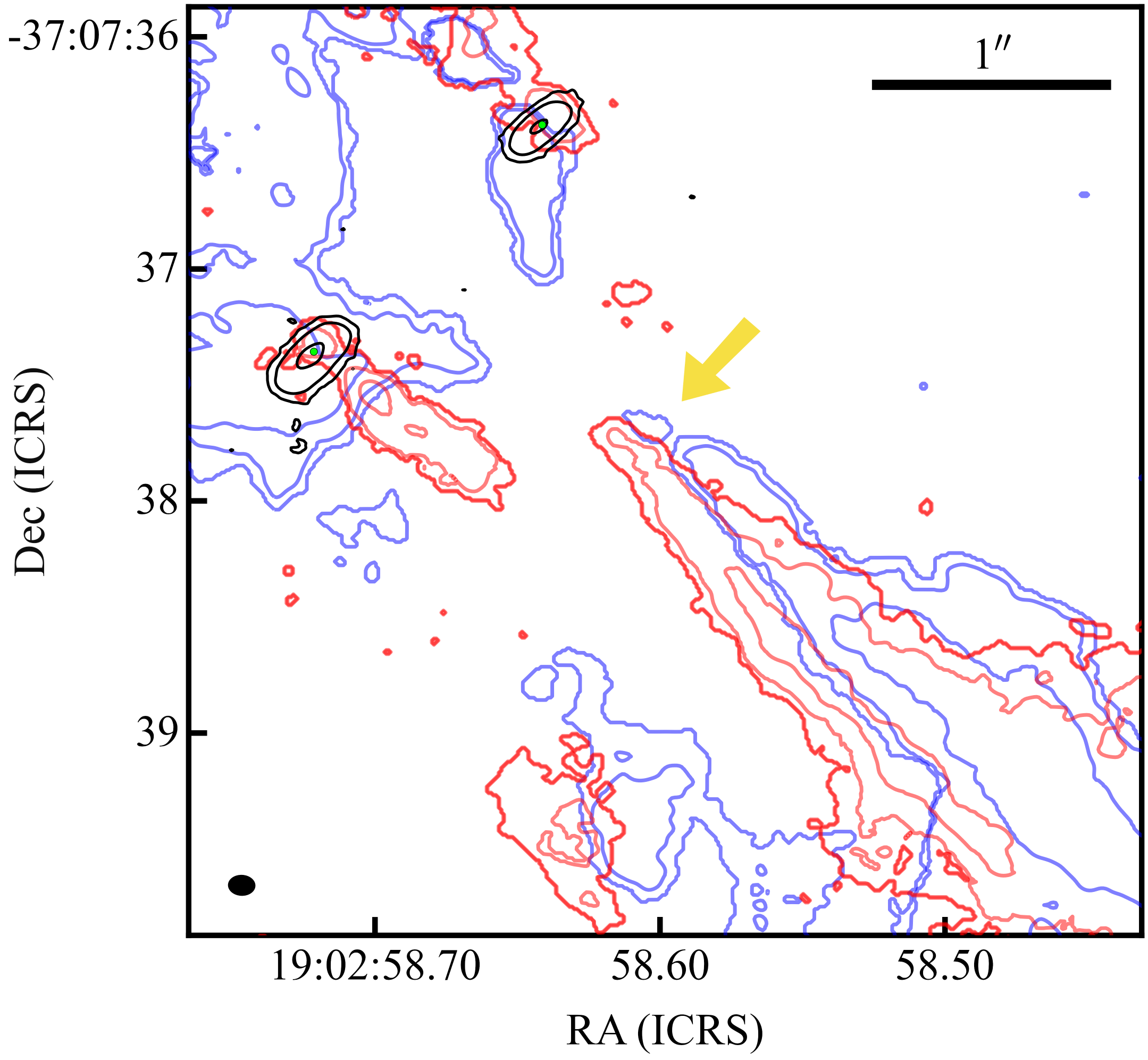}
  \caption{ IRAS 32 $^{12}$CO moment 0 map to illustrate the red/blue interaction region.  The blue contours correspond with -5.53 to -3.63 km s$^{-1}$ and the red contours correspond with 3.99 to 8.44 km s$^{-1}$. The black contours are the continuum emission. The yellow arrow points at the possible interaction region in the southwest region as referenced in \S 4.4. The beam size is 0$\farcs$11 $\times$ 0$\farcs$09 arcseconds with a PA of 87.2$^{\circ}$.}
  \label{fig:906e}
\end{figure}


As mentioned above, there are bright red and blue $^{12}$CO structures in the southwest highlighted in Fig. \ref{fig:906e}. The structures begin somewhat equidistant southwest of the protostars and extend to the southwest; the structures are detected in the higher velocity red and blue components. The blue structure is peaking generally to the north of the red structure (see Fig. \ref{fig:906e}). 
In addition, on the opposite side, to the northeast of the binary pair, there is a red component in channels with velocities of 2.09 and 2.72 km s$^{-1}$ that is best seen in Figure \ref{fig:906c}. There is also a blue component in channels with velocities of -8.07 to -4.06  km s$^{-1}$ but with much more contamination by the blue shifted lobe.  The northern interaction region is in general less defined compared to the south. 

\begin{figure}[h]
  \centering
    \includegraphics[angle=0,width=0.50\textwidth]{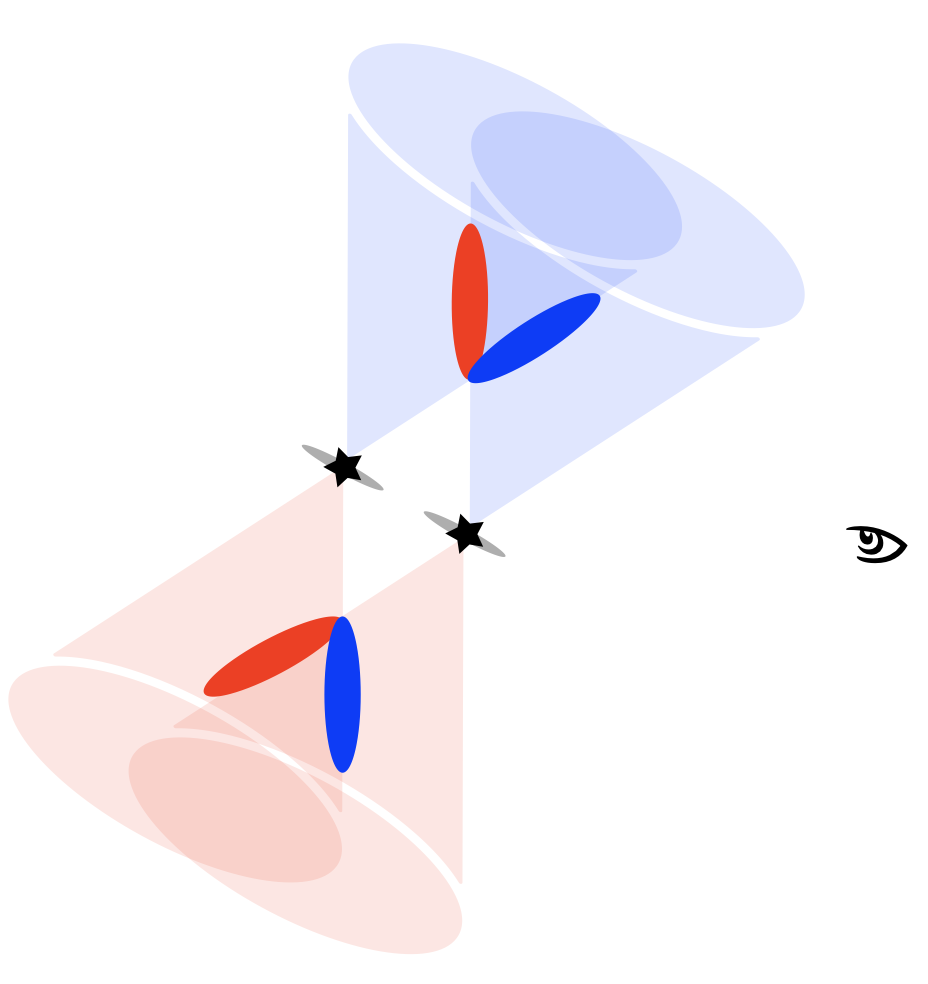}
  \caption{Simple cartoon of the IRAS 32 outflow interaction (side-view with the observer to the right) that may explain the red/blue outflow features seen in CO. The interaction is close to the plane of sky, so the interaction flow presents as both red and blue shifted gas in both lobes. Note, not to scale and the outflow lobes are colored red and blue based on the observer viewpoint.}
  \label{fig:920}
\end{figure}

The mostly likely explanation for these features   is two regions where the outflows driven by each of the binary disks A and B are interacting.  If we consider the two outflows as cones of entrained material then the two outflows will intercept each other, 
resulting in flows or possibly shocks that enhance the emission toward the observer and away from the observer (see Figure \ref{fig:920} for a cartoon of one possible configuration). On the south side of the binary, the interaction is more pronounced than on the north side. With highly inclined protostars, it is common for outflows to exhibit both redshifted and blueshifted emission on both sides of the outflow \citep[e.g., the nearly edge on eDisk source L1527,][]{vantHoff2023}, so with an inclination of around 70$^{\circ}$ in IRAS32, the interacting gas easily flows blueshifted and redshifted on both sides. In this case, the blue component and the red component are seen as nearly co-spatial along the line of sight. Note that Fig. \ref{fig:906e} also nicely shows red jet emission from Source A.

Although we do not detect an SiO jet in either outflow component of the binary, this is not uncommon in Class 0 sources as about only 33\% of the sources with an internal luminosity between 0.1 and 1 L$_\odot$ have detected SiO emission \citep{padio2021}; IRAS 32 has a total luminosity for both sources of 1.6 L$_\odot$, hence each source individually may be too weak to drive a detectable SiO jet.

\vspace{0.1cm}
\subsection{SO structure}

\begin{figure}[h]
  \centering
    \includegraphics[width=0.45\textwidth,trim=2 2 3 2,clip]{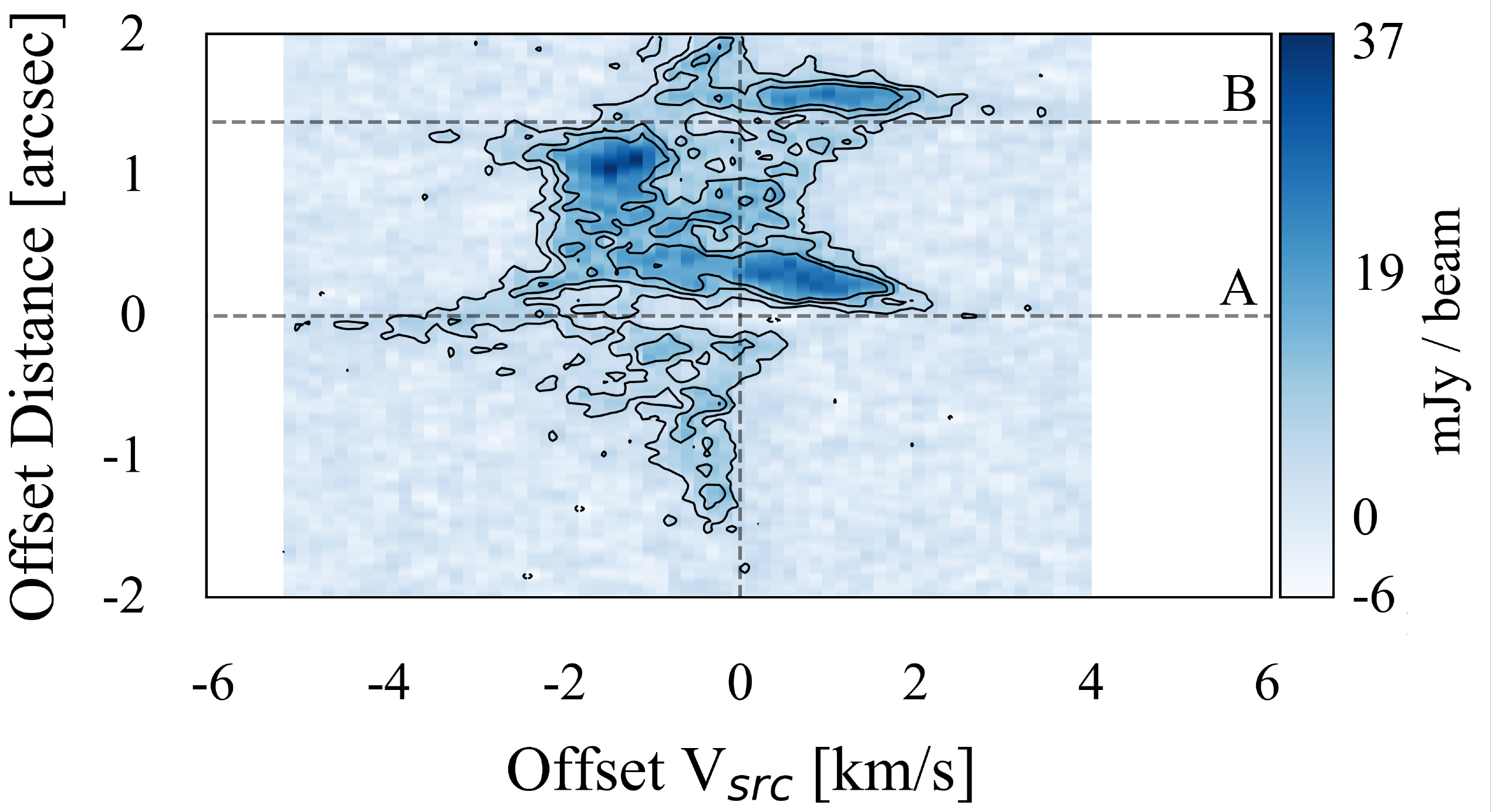}
  \caption{Same as Fig. \ref{fig:907b}, top (i.e. centered on source A), but with SO.
  The black contours are at 3$\sigma$n for $\sigma$=1.71 mJy beam$^{-1}$ and n$\in$[1,2,3]. The window size was chosen to be identical to Fig. \ref{fig:907b}.  Note that this PV diagram is for source A, using its Gaussian fit for the PA of this slice.  This means that the actually position of source B is slightly off slice center (much less than a beam) toward the north.
  }
  \label{fig:907d}
\end{figure}

The SO moment maps, in Figs. \ref{fig:904} and \ref{fig:904b}, show structures influenced by the Keplerian motion of the circumbinary disk and the circumstellar disks. 
The velocities in these features are not as large as the corresponding velocities in $^{13}$CO in Fig. \ref{fig:902}B, (e.g., also compare Fig.  \ref{fig:907d} with Fig. \ref{fig:907b}), suggesting that the SO is tracing circumstellar disk Keplerian material with generally larger radii.
As also seen in Fig. \ref{fig:904}, the filamentary structures (described in \S \ref{SOemission}) exhibit slight  velocity gradients toward the binaries. The filament kinematics in Fig. \ref{fig:904} show a blue component in the north and a red component in the south. We interpret these asymmetric structures as perhaps streamers between the envelope and the circumstellar and/or circumbinary disks.

With Fig. \ref{fig:907d}, we can compare the PV diagram of the SO emission to the $^{13}$CO emission in Fig. \ref{fig:907b}. Besides the lack of central velocities in $^{13}$CO, the two PV diagrams are broadly similar, although the SO velocity ranges are smaller for most features, as mentioned above. However, the velocity gradients in the PV diagrams are somewhat different in detail. On the blueshifted and redshifted sides of the disks in Fig. \ref{fig:907d}, one can again see hints of the Keplerian rotation (which is much more clear in the $^{13}$CO of Fig. \ref{fig:907b}).
One feature that is more noticeable in the SO PV diagram is the bright blue emission between the two disks, just below the blueshifted side of disk B in Fig. \ref{fig:907d}.  This is the location of the filament structure in SO in the northeast, suggesting a possible interaction from the incoming flow of material at that location. 

When comparing the SO and $c$-C$_3$H$_2$ line emission of IRAS 32 with those of the eDisk source L1527 \citep{vantHoff2023}, they have similar features. As discussed above, the SO line emission is also extended and traces the disk \citep[perhaps the disk surface layers, see][]{vantHoff2023} in both sources. The $c$-C$_3$H$_2$ in both sources is not detected in the outflow cavity but rather in the inner envelope (or circumbinary region). In fact, out of the eDisk survey IRAS 32 most resembles the detection of these two molecules with L1527.  It is possible that IRAS 32, like L1527, may be another example of a source that exhibits 
warm carbon-chain chemistry pathways \citep[eg.,][]{sakai2008}, wherein carbon-chain molecules are produced via CH$_4$ evaporated from dust grains.

\section{Conclusion} \label{sec:conclusion}

We observed the emission towards IRAS 32 at a resolution of $\sim$0\farcs03 in the 1.3 mm dust continuum and $\sim$0\farcs1 for numerous molecular lines, including  $^{12}$CO ($J=2\rightarrow1$), $^{13}$CO ($J=2\rightarrow1$), C$^{18}$O ($J=2\rightarrow1$), and SO ($J=6_5\rightarrow5_4$), as a part of the ALMA eDisk Large Program. Our conclusions are:

\begin{enumerate}

    \item IRAS 32 is a Class 0 newly discovered binary protostellar system with circumstellar disks surrounding both protostars and a  circumbinary disk. The circumstellar disks are aligned parallel to each other and to the (projected) orbital plane, indicating that this particular binary system is likely  formed in a relatively ordered fashion, such as disk fragmentation (as opposed to turbulent fragmentation followed by orbital migration). 
    
    \item We detected all the molecular lines in Table \ref{tab:spec_data} except for SiO, CH$_{3}$OH, and the 218.16 transition of $c$-C$_3$H$_2$. Maps of CO, $^{13}$CO, and C$^{18}$O show outflows in the NE-SW direction, normal to the circumbinary disk, and the aligned circumstellar disks. SO emission mainly shows the inner envelope/circumbinary disk and the circumstellar disks structures. The remaining molecular lines were detected in only a few channels and seemed to mostly favor the SW region of the binary system.
    
    \item IRAS 32 A has a Gaussian-fit estimated disk radius of 180$\pm$2 millarcseconds (26.9$\pm$0.3 au), whereas the IRAS 32 B disk is estimated at 153$\pm$2 millarcseconds (22.8$\pm$0.3 au). The circumbinary disk-like structure size is estimated at 3$\farcs$25 (488 au).

    \item The (northeast) far-side of the disk is brighter than the (southeast) near-side in dust continuum along the minor axis, which indicates that the dust emission is optically thick and the grains responsible for the continuum emission has yet to settle into a thin layer near the mid-plane. 

    \item No clear substructures are detected in the circumstellar disks. The radial profile of the circumstellar disks are non-Gaussian. More careful modeling is required.
    
    \item The dust mass for protostars A and B are given as either a) 21.9$\pm$1.1 and 12.7$\pm$0.6 M$_{\Earth}$ for an average temperature derived from a radiative transfer calculation using the bolometric luminosity as prescribed by \cite{tobin2020}, or b) 52.0$\pm$2.6 and 29.1$\pm$1.4 M$_{\Earth}$ for T=20 K.
    
    \item The protostar masses, derived from SLAM fits to the PV diagrams of $^{13}$CO ($J=2\rightarrow1$) emission in the circumstellar disks, are estimated to be 0.91$\pm$0.01 M$_\odot$ and 0.32$\pm$0.01 M$_\odot$ for the edge fit and 0.50$\pm$0.55 M$_\odot$ and 0.48$\pm$0.03 M$_\odot$ for the ridge fit, for source A and B, respectively. The errors here are large due to the complicated nature of the binary system. Since the stellar masses are likely somewhere between the edge and ridge fits, we find expected stellar masses of $<$0.92 M$_\odot$ and 0.31-0.51 M$_\odot$, for sources A and B, respectively.


\end{enumerate}


\acknowledgments
FJE and LWL acknowledge support from NSF AST-2108794. ST is supported by JSPS KAKENHI grant Nos. 21H00048 and 21H04495, and by NAOJ ALMA Scientific Research grant No. 2022-20A.
J.J.T. acknowledges support from NASA XRP 80NSSC22K1159.
J.K.J. acknowledges support from the Independent Research Fund Denmark (grant No. 0135-00123B). Y.A. acknowledges support by NAOJ ALMA Scientific Research Grant code 2019-13B, Grant-in-Aid for Scientific Research (S) 18H05222, and Grant-in-Aid for Transformative Research Areas (A) 20H05844 and 20H05847.
W.K. was supported by the National Research Foundation of Korea (NRF) grant funded by the Korea government (MSIT) (NRF-2021R1F1A1061794).
N.O. acknowledges support from National Science and Technology Council (NSTC) in Taiwan through grants NSTC 109-2112-M-001-051 and 110-2112-M-001-031.
Z.-Y.L. is supported in part by NASA NSSC20K0533 and NSF AST-2307199 and AST-1910106.
ZYDL acknowledges support from NASA 80NSSCK1095, the Jefferson Scholars Foundation, the NRAO ALMA Student Observing Support (SOS) SOSPA8-003, the Achievements Rewards for College Scientists (ARCS) Foundation Washington Chapter, the Virginia Space Grant Consortium (VSGC), and UVA research computing (RIVANNA). IdG acknowledges support from grant PID2020-114461GB-I00, funded by MCIN/AEI/10.13039/501100011033. H.-W.Y. acknowledges support from National Science and Technology Council (NSTC) in Taiwan through grant NSTC 110-2628-M-001-003-MY3 and from the Academia Sinica Career Development Award (AS-CDA-111-M03).

This paper makes use of the following ALMA data: ADS/JAO.ALMA\#2019.1.00261.L. ALMA is a partnership of ESO (representing its member states), NSF (USA) and NINS (Japan), together with NRC (Canada), MOST and ASIAA (Taiwan), and KASI (Republic of Korea), in cooperation with the Republic of Chile. The Joint ALMA Observatory is operated by ESO, AUI/NRAO and NAOJ. The National Radio Astronomy Observatory is a facility of the National Science Foundation operated under cooperative agreement by Associated Universities, Inc.

The National Radio Astronomy Observatory is a facility of the National Science Foundation operated under cooperative agreement by Associated Universities, Inc.

\clearpage
\appendix
\label{sec:appendix}

\section{Moment 0 and Channel Maps}

\begin{figure*}[h]
  \centering
    \includegraphics[width=1.0\textwidth]{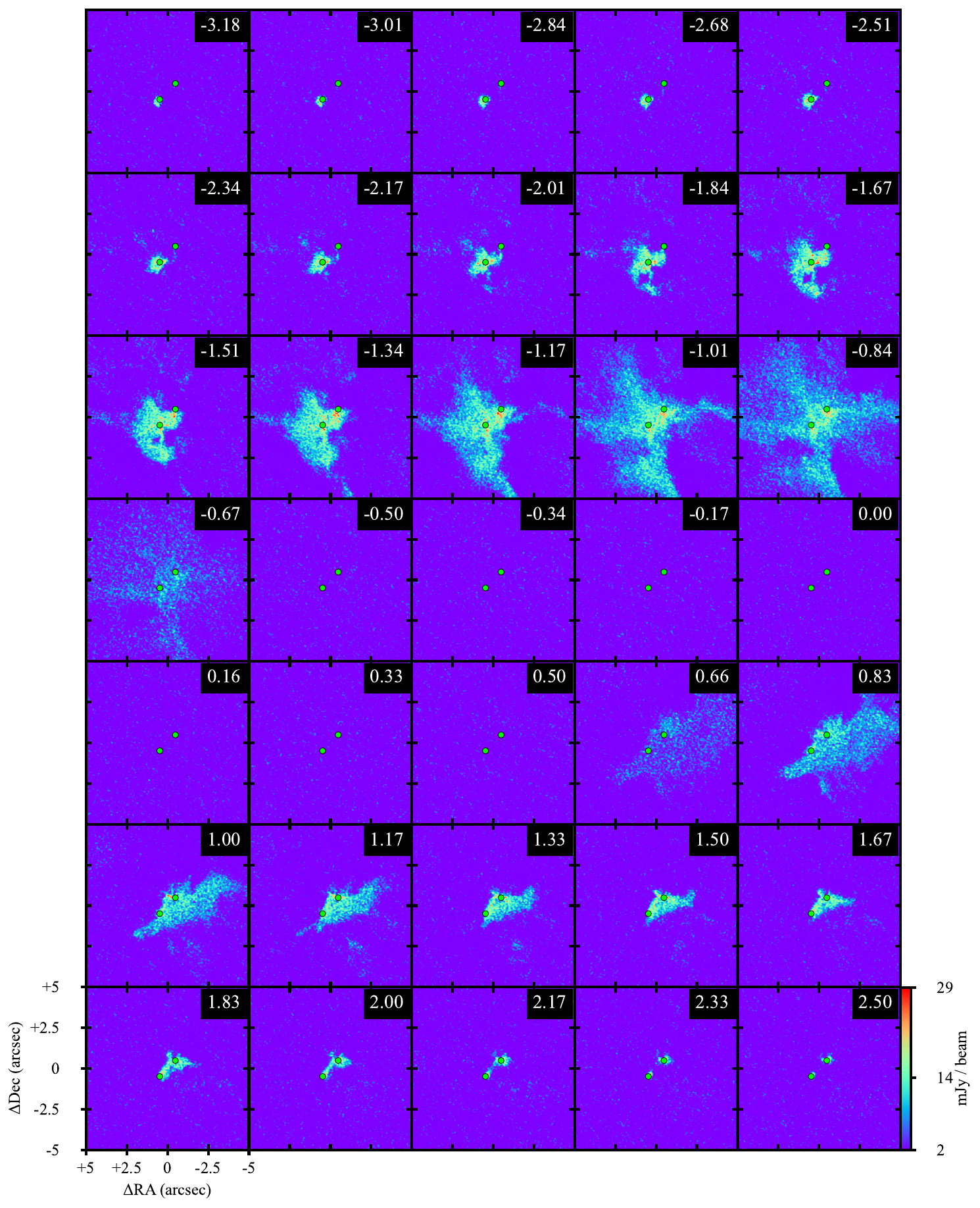}
  \caption{Same as Fig. \ref{fig:906c} but for $^{13}$CO. The beam size is 0$\farcs$11 $\times$ 0$\farcs$09 with a PA of 84.2$^{\circ}$.}
  \label{fig:908h}
\end{figure*}

\begin{figure*}[h]
  \centering
    \includegraphics[width=1.0\textwidth]{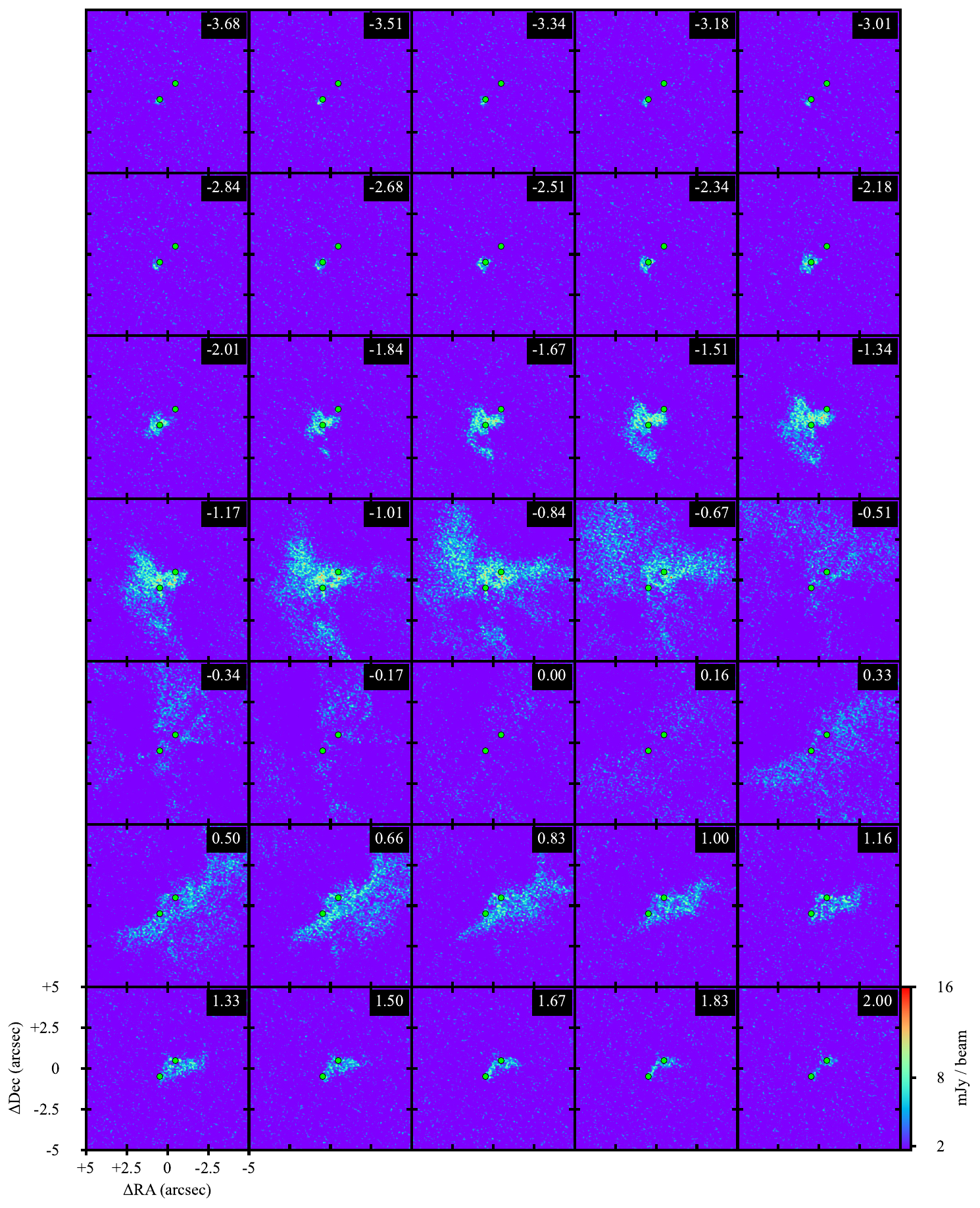}
  \caption{Same as Fig. \ref{fig:906c} but for C$^{18}$O. The beam size is 0$\farcs$11 $\times$ 0$\farcs$09 with a PA of 84.9$^{\circ}$.}
  \label{fig:908i}
\end{figure*}

\begin{figure*}[h]
  \centering
    \includegraphics[width=1.0\textwidth]{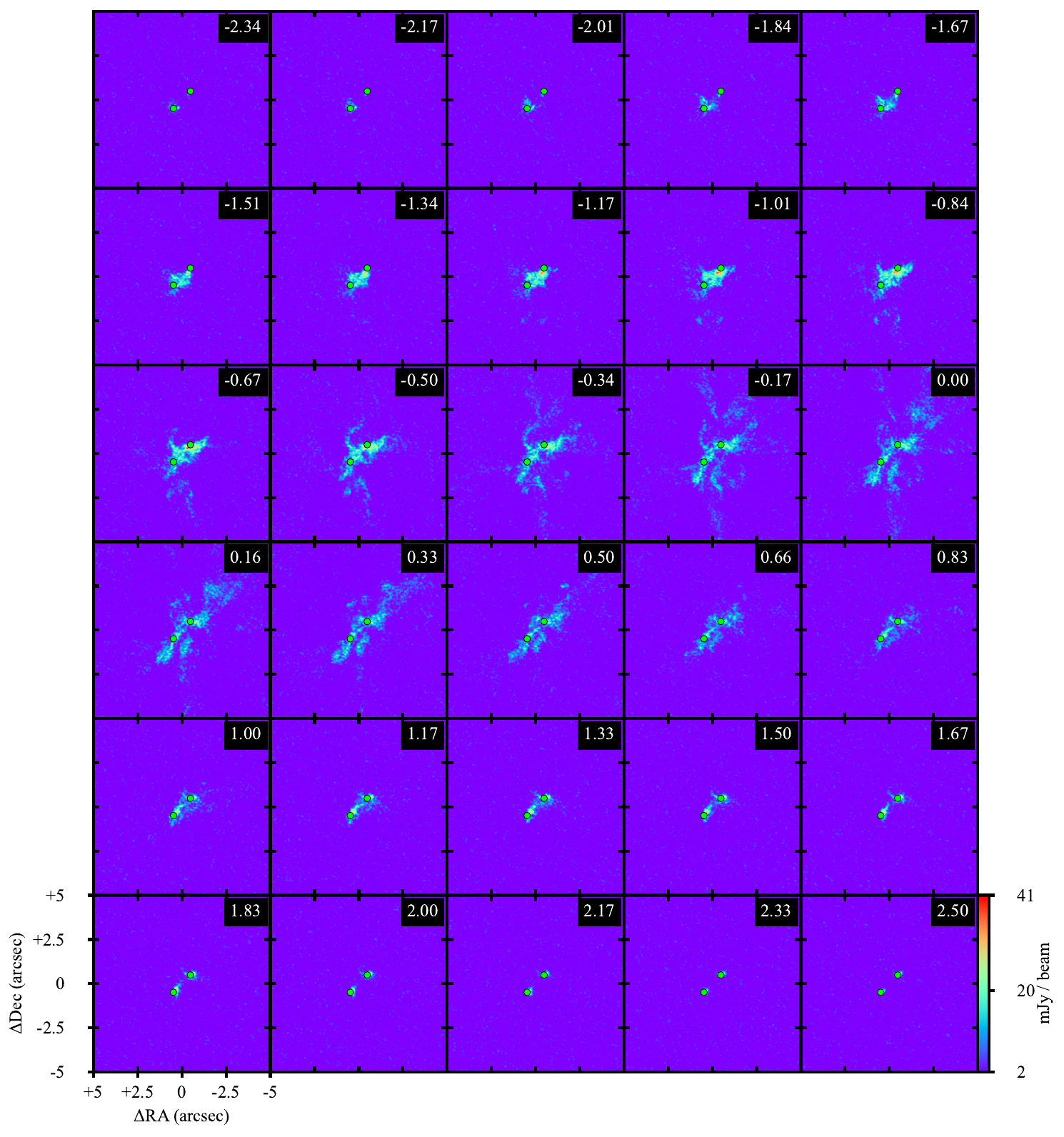}
  \caption{Same as Fig. \ref{fig:906c} but for SO. The beam size is 0$\farcs$11 $\times$ 0$\farcs$09 with a PA of 85.1$^{\circ}$.}
  \label{fig:908j}
\end{figure*}

This section includes the moment 0 and channel maps for all the molecular lines with sufficient emission that were not already shown in the main text. This includes the channel maps of $^{13}$CO, C$^{18}$O, and SO in Fig. \ref{fig:908h}, \ref{fig:908i}, and \ref{fig:908j}, respectively. 
All of the lines are consistent with the analysis presented in the main text. 

\subsection{DCN}

\begin{figure*}[h]
  \centering
    \includegraphics[width=0.5\textwidth]{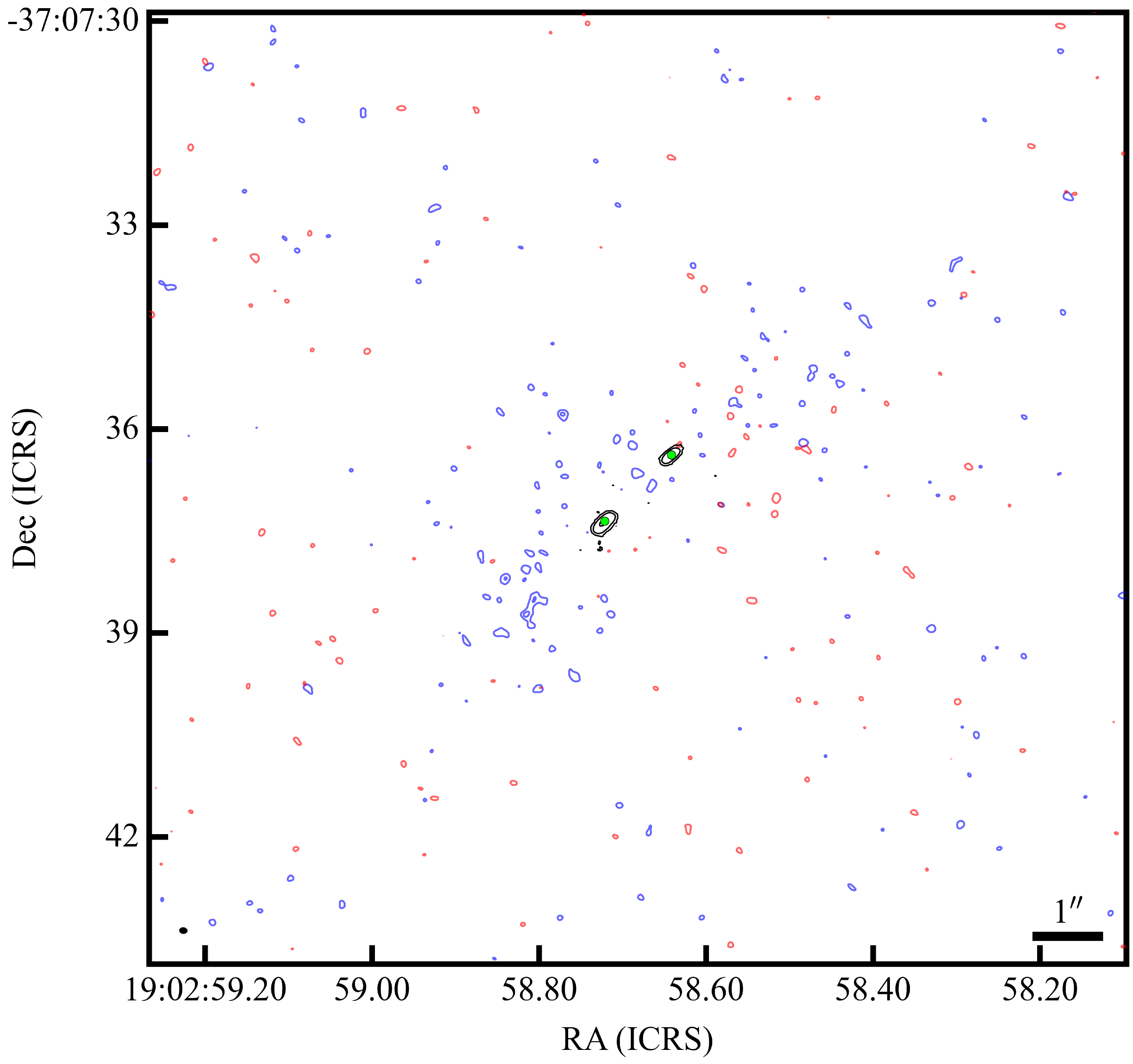}
  \caption{IRAS 32 DCN (3-2) moment 0 maps using robust = 0.5 with both short and long baseline data {\bf were made using the velocity range from Table \ref{tab:spec_results}.} The blue and red contours are at 2$\sigma$ and 3$\sigma$, which is defined as $\sigma_{blue}$ = 1.35 and $\sigma_{red}$ = 1.32 mJy km s$^{-1}$. The Gaussian-fit centers are shown as lime circles. The beam size is 0$\farcs$11 $\times$ 0$\farcs$08 with a PA of 85$^{\circ}$.}
  \label{fig:908d2}
\end{figure*}
\begin{figure*}[h]
  \centering
    \includegraphics[width=0.83\textwidth]{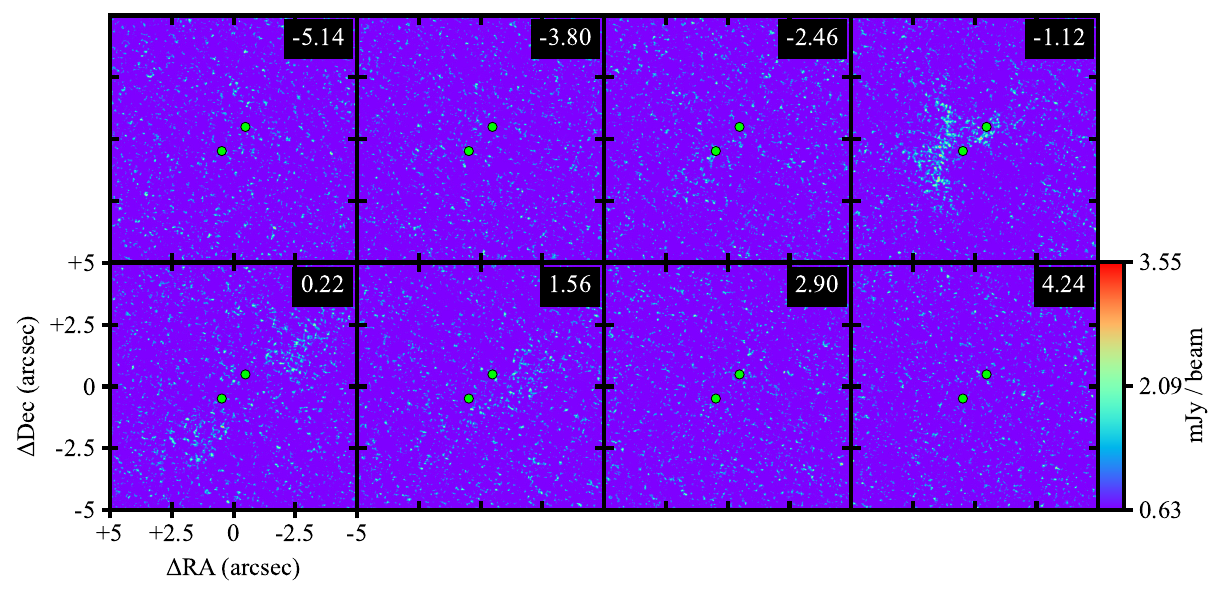}
  \caption{IRAS 32 DCN channel map using robust = 0.5 with both short and long baseline data combined. The Gaussian-fit centers are shown as lime circles. The velocity offset from 5.86 km s$^{-1}$ is shown in the top right of each channel. Each individual image is within a 10 by 10$^{\prime\prime}$ box. The beam size is 0$\farcs$11 $\times$ 0$\farcs$08 with a PA of 85$^{\circ}$.}
  \label{fig:908d}
\end{figure*}

The DCN (3-2) emission is shown in Figs. \ref{fig:908d2} and \ref{fig:908d}.
With an adopted rest velocity at 5.86 km s$^{-1}$, DCN has emission in about 4 channels (-2.46, -1.12, 0.22, and 1.56 km s$^{-1}$), which slightly favors the blue north-east. Morphologically, the blue and red sides look most similar to C$^{18}$O in Fig. \ref{fig:908i}, which traces a combination of circumbinary disk and outflow although the blue side of DCN seems to trace out more of the NE outflow cavity than the C$^{18}$O.

\subsection{\texorpdfstring{$c$-C$_3$H$_2$}{c-C3H2}}

\begin{figure*}[h]
  \centering
    \includegraphics[width=0.5\textwidth]{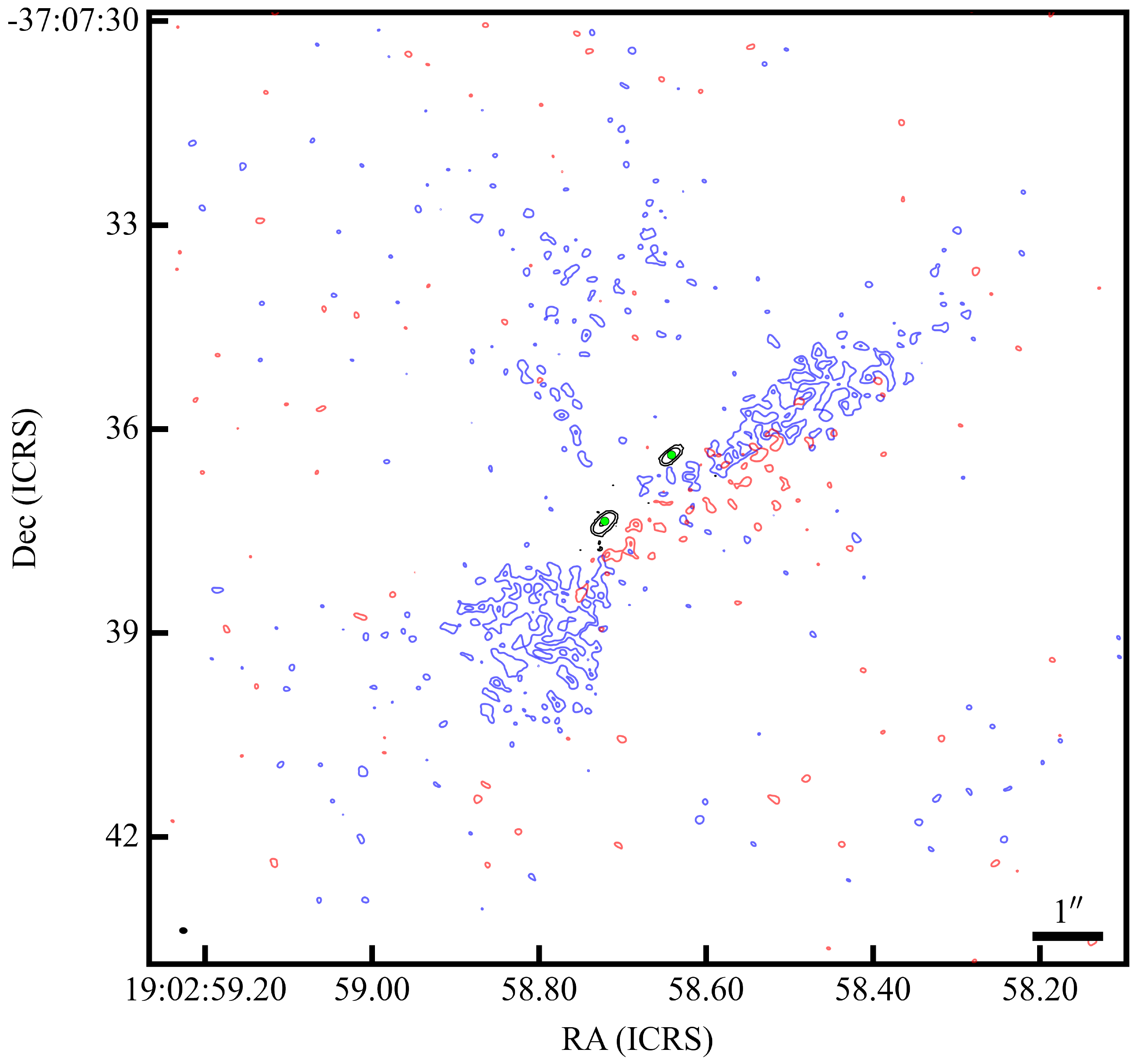}
  \caption{
  Same as Fig. \ref{fig:908d2} but for the blended c-C$_3$H$_2$ 217.82 GHz (6$_{0,6}$-5$_{1,5}$) and (6$_{1,6}$-5$_{0,5}$) lines with $\sigma_{blue}$ = 1.26 and $\sigma_{red}$ = 1.26 mJy km s$^{-1}$.}
  \label{fig:908a2}
\end{figure*}
\begin{figure*}[h]
  \centering
    \includegraphics[width=0.83\textwidth]{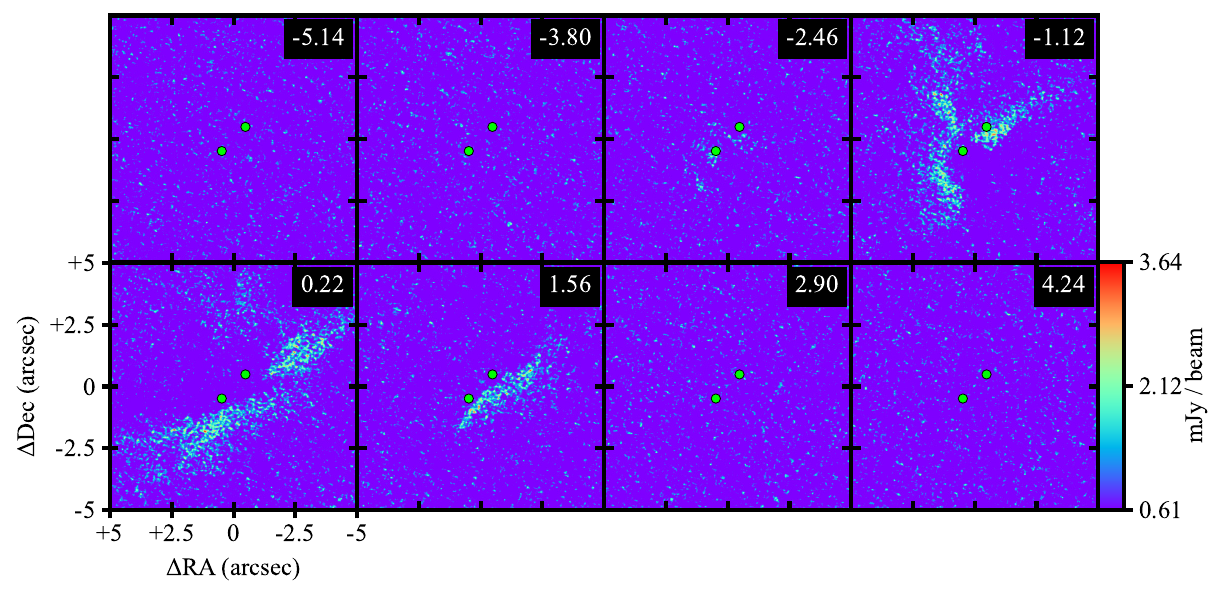}
  \caption{Same as Fig. \ref{fig:908d} but for c-C$_3$H$_2$ 217.82 GHz.}
  \label{fig:908a}
\end{figure*}

\begin{figure*}[h]
  \centering
    \includegraphics[width=0.5\textwidth]{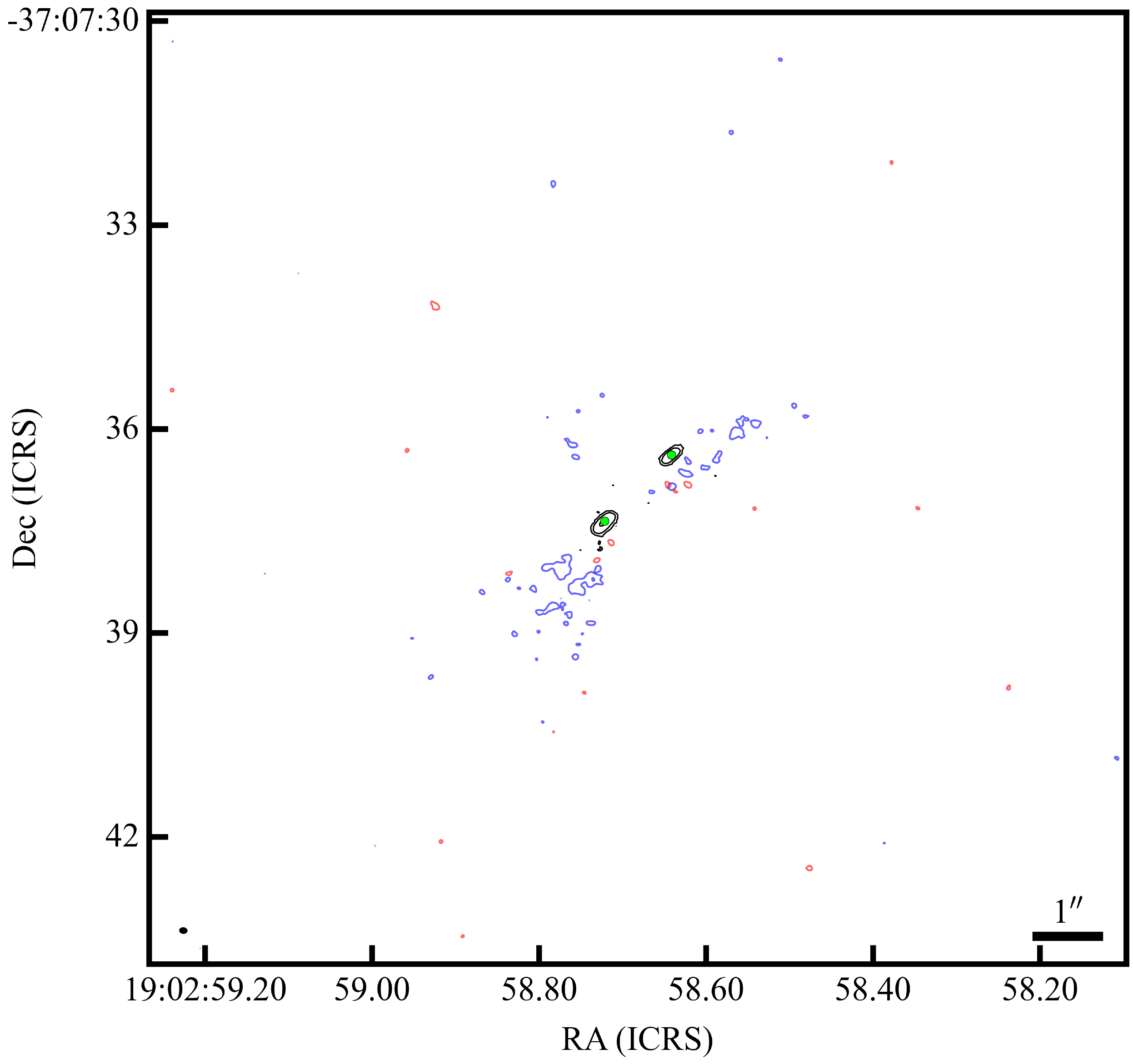}
  \caption{
  Same as Fig. \ref{fig:908d2} but for c-C$_3$H$_2$ 217.94 GHz (5$_{1,4}$-4$_{2,3}$) with $\sigma_{blue}$ = 1.20 and $\sigma_{red}$ = 1.16 mJy km s$^{-1}$.}
  \label{fig:908b2}
\end{figure*}

\begin{figure*}[h]
  \centering
    \includegraphics[width=0.83\textwidth]{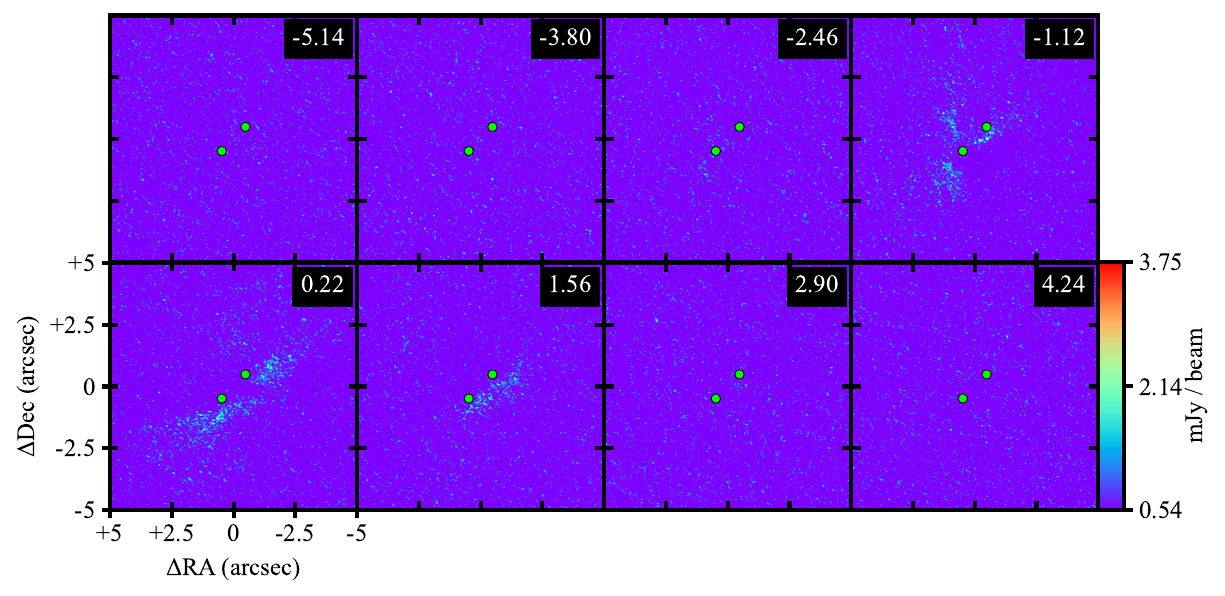}
  \caption{Same as Fig. \ref{fig:908d} but for c-C$_3$H$_2$ 217.94 GHz.}
  \label{fig:908b}
\end{figure*}

The blended line of c-C$_{3}$H$_{2}$ at 217.82 GHz (6$_{0,6}$-5$_{1,5}$) and (6$_{1,6}$-5$_{0,5}$) is shown in Figs. \ref{fig:908a2} and \ref{fig:908a}, which also has emission in the same 4 channels as DCN (-2.46, -1.12, 0.22, and 1.56 km s$^{-1}$). The c-C$_{3}$H$_{2}$ is mostly tracing a flattened region surrounding the binaries.  This morphology is most similar to the $^{13}$CO and C$^{18}$O lines in Figs. \ref{fig:908h} and \ref{fig:908i}, respectively. Although c-C$_{3}$H$_{2}$ is often considered a good tracer of irradiated gas inside outflow cavities, in this case, c-C$_{3}$H$_{2}$ is mostly tracing the circumbinary disk or inner envelope. This can be compared to eDisk source L1527 \citep{vantHoff2023}, where the c-C$_{3}$H$_{2}$ emission is also tracing a flattened region of the inner envelope.
The 217.94 GHz transition (5$_{1,4}$-4$_{2,3}$) is shown in Figs. \ref{fig:908b2} and \ref{fig:908b} with a very similar  emission, albeit with weaker emission.






\subsection{\texorpdfstring{H$_2$CO}{H2CO}}

\begin{figure*}[h]
  \centering
    \includegraphics[width=0.5\textwidth]{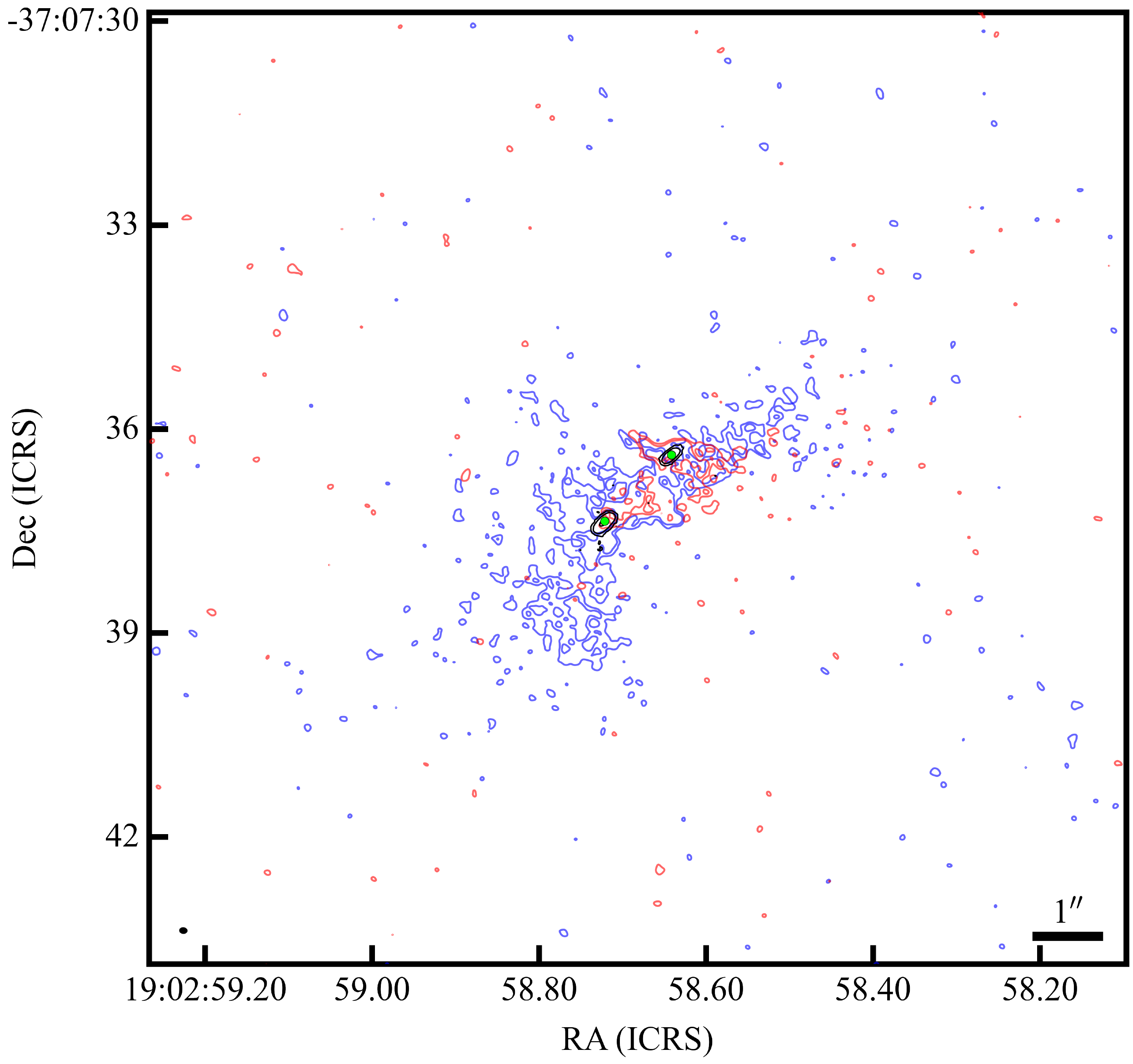}
  \caption{Same as Fig. \ref{fig:908d2} but for H$_{2}$CO 218.22 GHz (3$_{0,3}$-2$_{0,2}$) with a velocity range from Table \ref{tab:spec_results},
   $\sigma_{blue}$ = 1.50 and $\sigma_{red}$ = 1.46 mJy km s$^{-1}$.}
  \label{fig:908e2}
\end{figure*}

\begin{figure*}[h]
  \centering
    \includegraphics[width=0.83\textwidth]{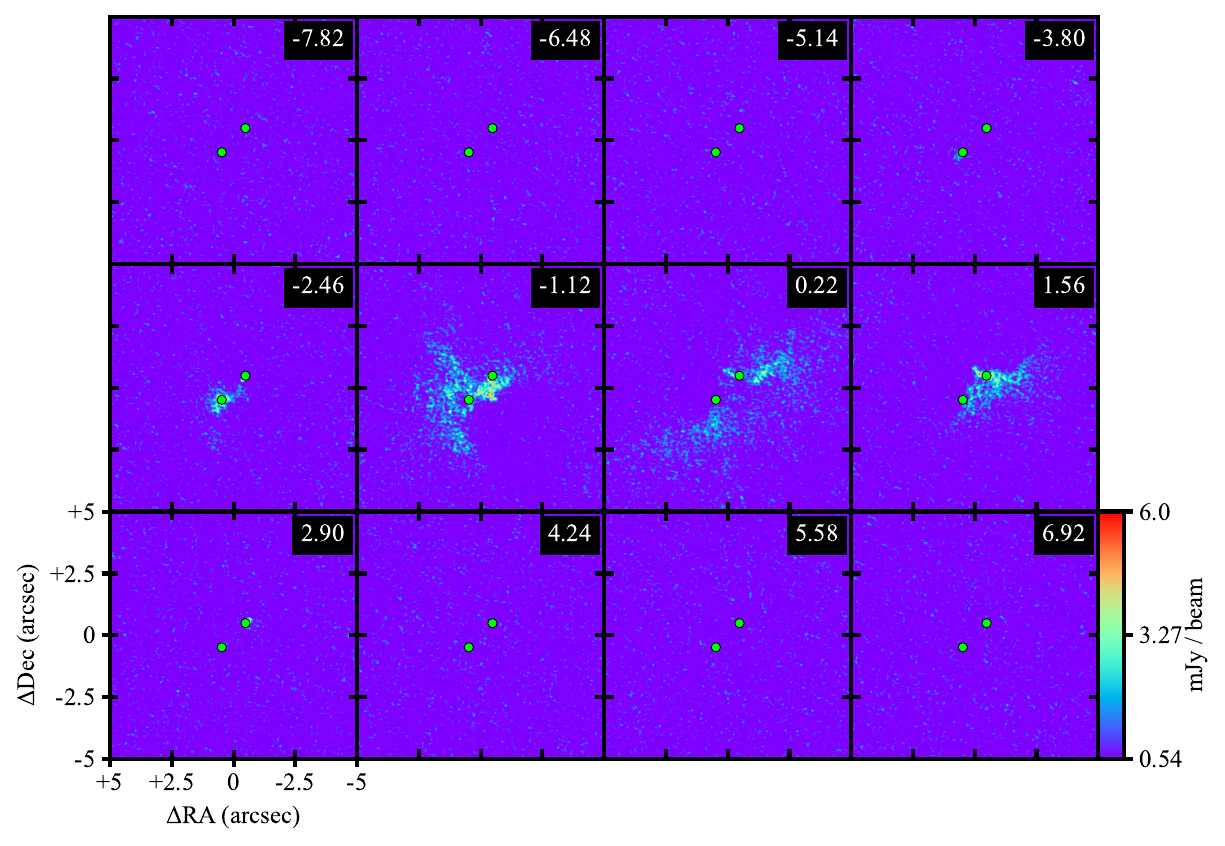}
  \caption{Same as Fig. \ref{fig:908d} but for H$_{2}$CO 218.22 GHz.}
  \label{fig:908e}
\end{figure*}

\begin{figure*}[h]
  \centering
    \includegraphics[width=0.5\textwidth]{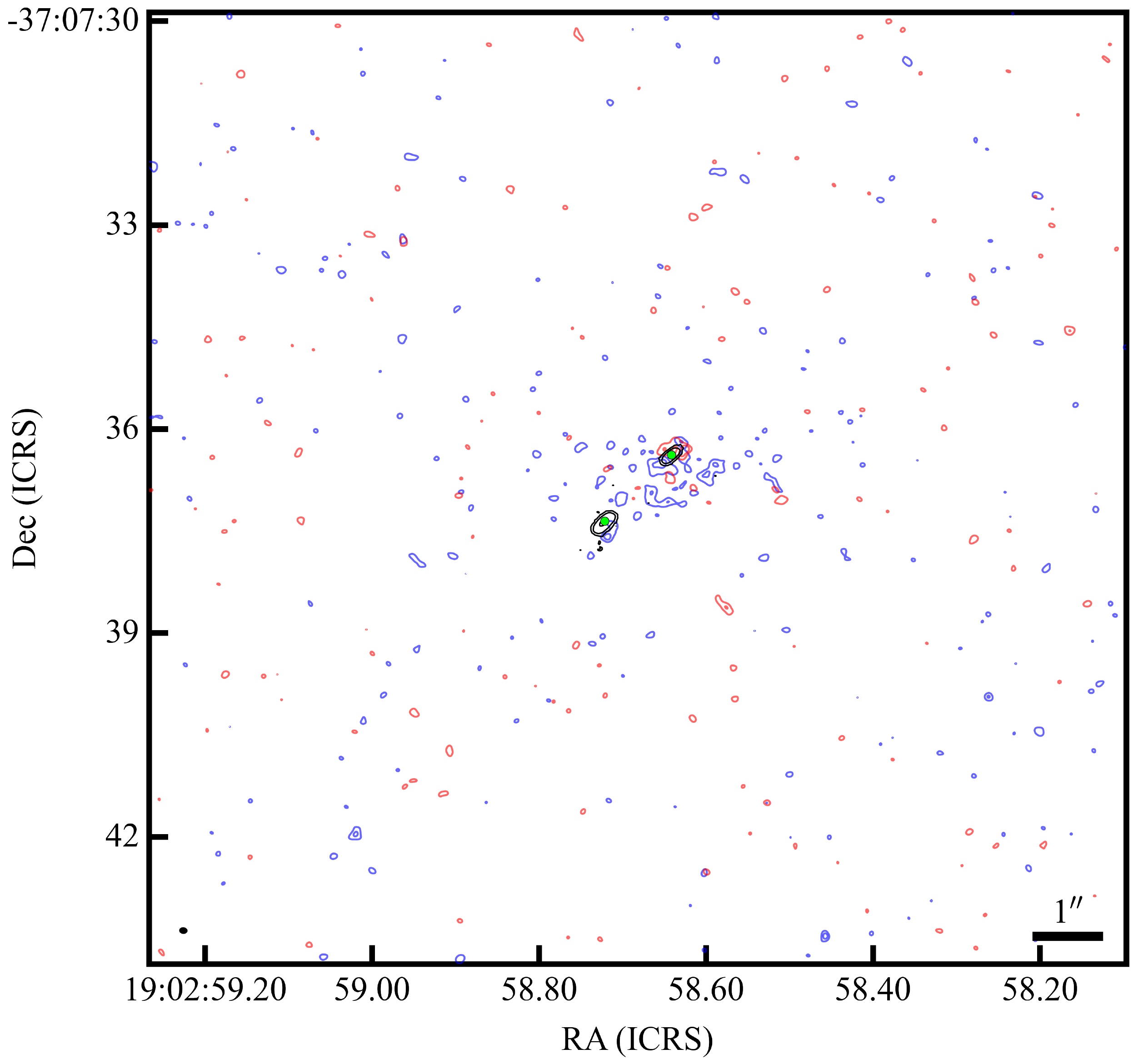}
  \caption{
  Same as Fig. \ref{fig:908d2} but for H$_{2}$CO 218.48 GHz (3$_{2,2}$-2$_{2,1}$) with $\sigma_{blue}$ = 1.14 and $\sigma_{red}$ = 1.47 mJy km s$^{-1}$.}
  \label{fig:908f2}
\end{figure*}

\begin{figure*}[h]
  \centering
    \includegraphics[width=0.83\textwidth]{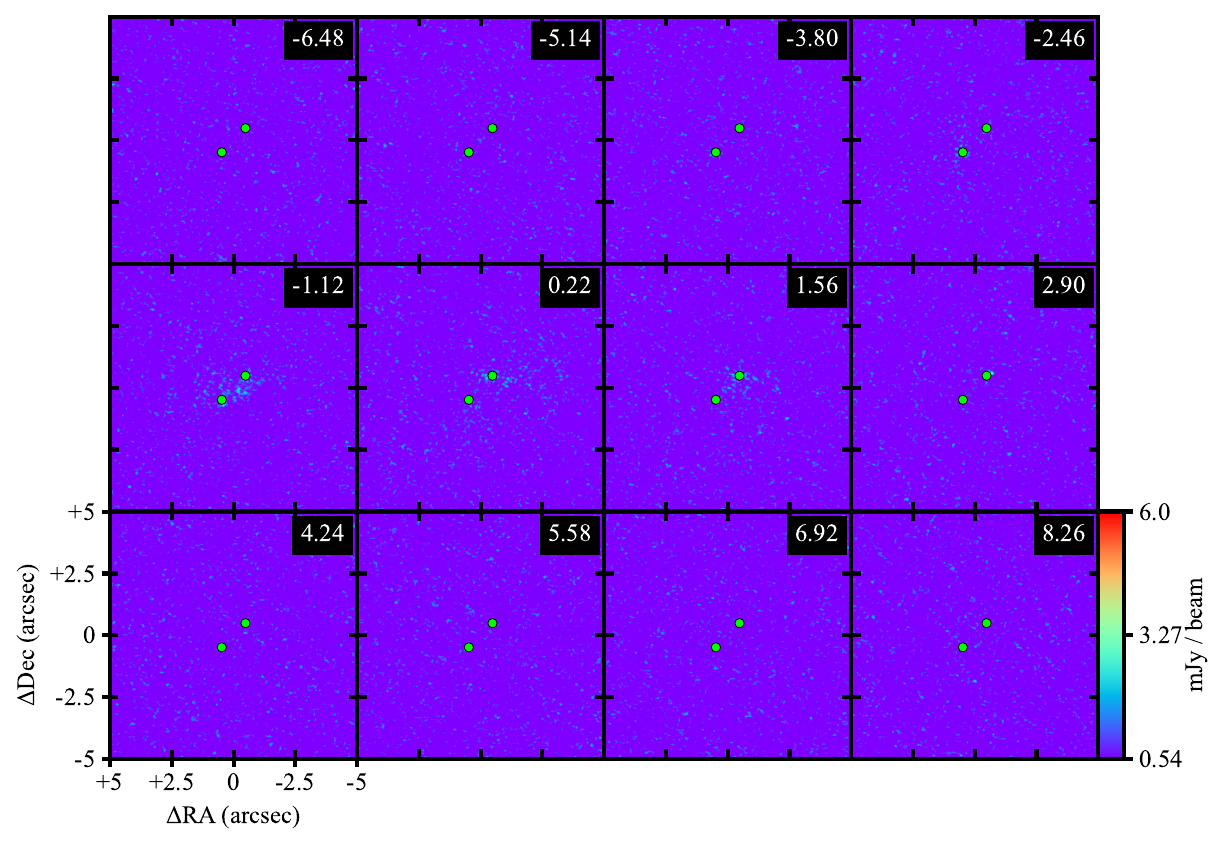}
  \caption{Same as Fig. \ref{fig:908d} but for H$_{2}$CO 218.48 GHz.}
  \label{fig:908f}
\end{figure*}

\begin{figure*}[h]
  \centering
    \includegraphics[width=0.5\textwidth]{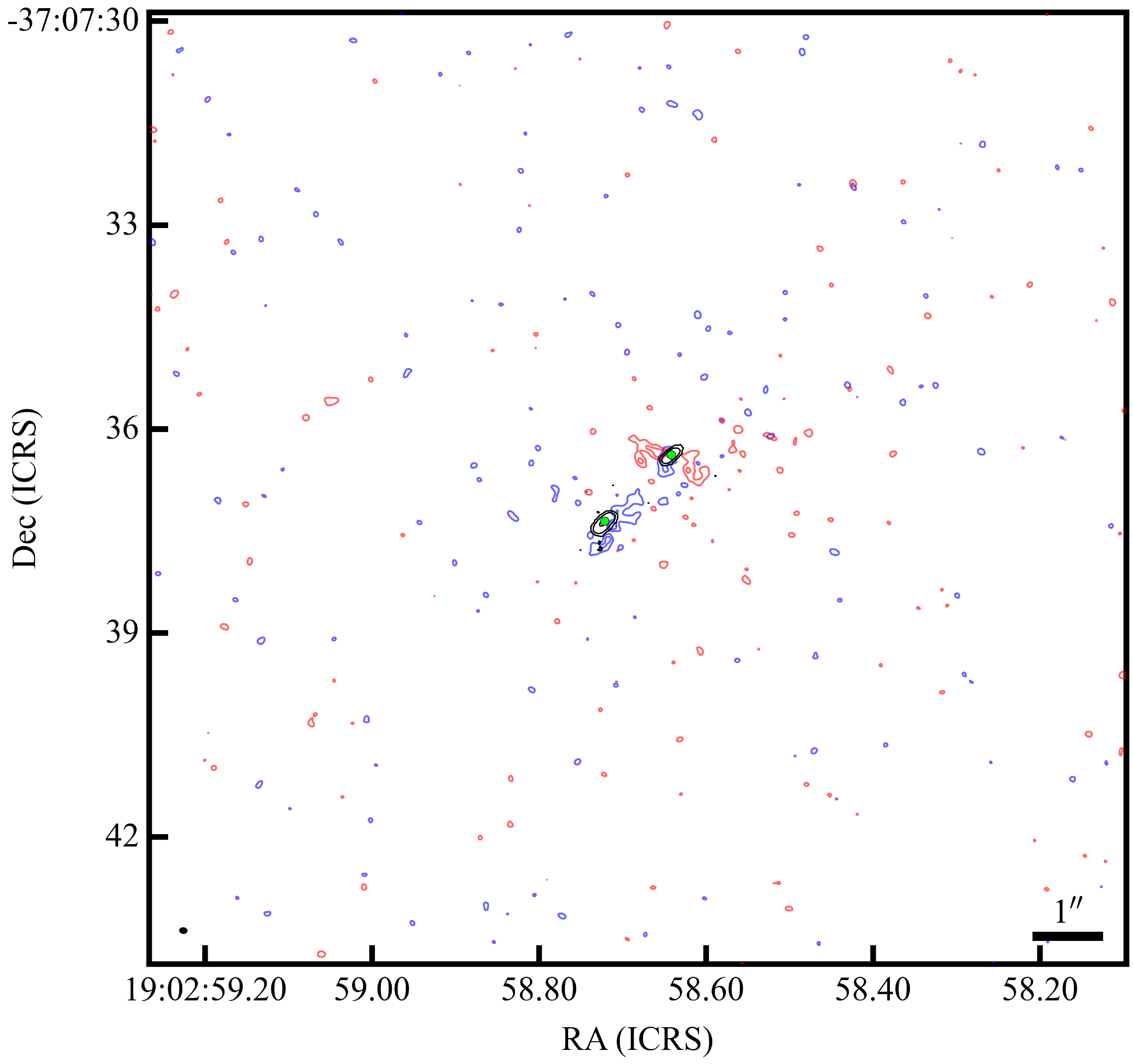}
  \caption{
  Same as Fig. \ref{fig:908d2} but for H$_{2}$CO 218.76 GHz (3$_{2,1}$-2$_{2,0}$) with $\sigma_{blue}$ = 1.26 and $\sigma_{red}$ = 0.96 mJy km s$^{-1}$.}
  \label{fig:908g2}
\end{figure*}

\begin{figure*}[h]
  \centering
    \includegraphics[width=0.83\textwidth]{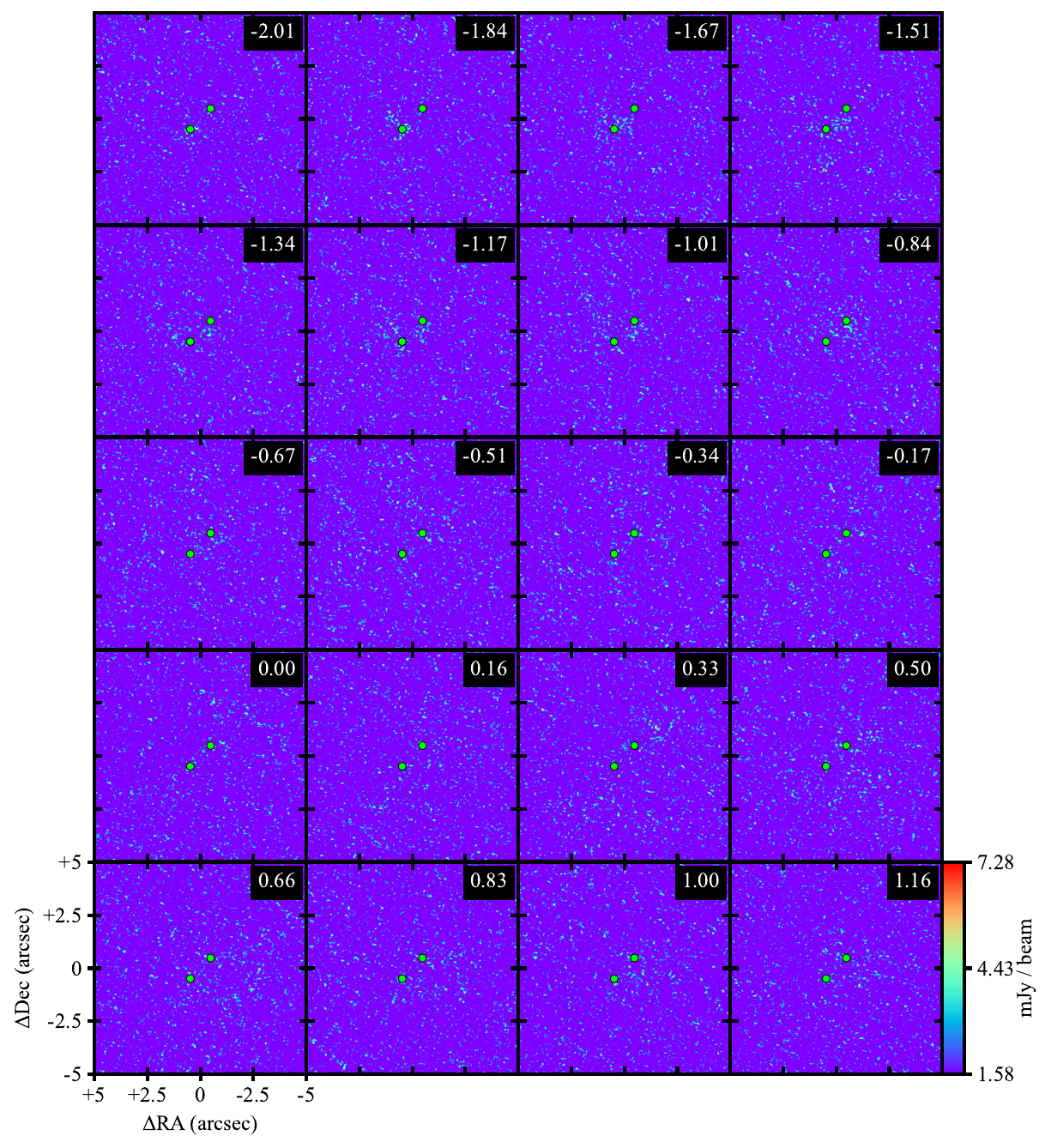}
  \caption{Same as Fig. \ref{fig:908d} but for H$_{2}$CO 218.76 GHz.}
  \label{fig:908g}
\end{figure*}

Formaldehyde at 218.22 GHz (3$_{0,3}$-2$_{0,2}$) is shown in Figs. \ref{fig:908e2} and \ref{fig:908e}.  It has emission from -3.80 to 2.90 km s$^{-1}$ in steps of $\Delta$v = 1.34 km s$^{-1}$. The relatively high velocity channels show emission around the SE side (blue) of protostar A and the NW side (red) of protostar B. As we approach the lower velocity channels, the emission begins to look like the $^{13}$CO map in Fig. \ref{fig:908h} with the blue side predominately in the SE region and the NE outflows, and the red side predominately in the SW side of the binary. However, there is a strong emission feature on the E side of protostar B in the red channel. The only other equivalent feature is in $^{12}$CO. It is possible that this is the NE outflow of protostar B. Similarly, on the blue side there is a bright emission feature south of protostar B that is projected from either protostar. It does not seem to match with any other feature seen in the other molecular lines, but with much lower spectral resolution than the CO and SO lines, it can be difficult to compare.
The formaldehyde 218.48 GHz transition (3$_{2,2}$-2$_{2,1}$) is shown in Figs.  \ref{fig:908f2} and  \ref{fig:908f}. It is
similar except with weaker emission, the outer channels fading into the noise, and the sharp emission features at low velocities are subdued.
The formaldehyde 218.76 GHz transition (3$_{2,1}$-2$_{2,0}$) is shown in Figs. 
\ref{fig:908g2} and \ref{fig:908g}. It is
again similar to the other H$_2$CO lines with even weaker emission but with higher velocity resolution.















\bibliography{8_bib}{}
\bibliographystyle{aasjournal}

\end{document}